\documentclass[
aps,
pra,            
twocolumn,      
letterpaper,           
preprintnumbers,
amsmath,        
amssymb,        
floatfix]{revtex4-2}

\usepackage[T1]{fontenc}
\usepackage{graphicx}    
\usepackage{color}      
\usepackage{bm}         
\usepackage{braket}
\usepackage{subcaption}
\usepackage[export]{adjustbox}
\usepackage{placeins}
\usepackage{placeins}

\begin{document}
\title{\textbf{Quantum discord of Gaussian states in non-inertial frames}} 

 \author{Siwei Li}
 \author{Xiaofen Huang} 
 \email{huangxf1206@163.com}
 \affiliation{School of Mathematics and Statistics, Hainan Normal University, Haikou 571158, China}
 
\begin{abstract}
	We investigate the redistribution of continuous-variable Gaussian quantum discord under the Unruh effect for both one and two uniformly accelerated observers. The discord shared between the inertial mode and the accessible accelerated mode, as well as that between the two accessible accelerated modes, decays monotonically with acceleration and vanishes in the limit $a\to\infty$, while the Unruh effect generates discord in all causally disconnected Rindler mode pairs. The initial quantum correlation is therefore redistributed across Rindler partitions rather than destroyed. We further characterize how the squeezing parameter $s$ and field frequency $\omega$ control the discord in different mode-pair classes. For causally connected pairs, $s$ and $\omega$ exert comparable, interchangeable effects; for cross-region pairs, the squeezing parameter dominates; and for intra-observer cross-region pairs, the field frequency is more influential. In the two-observer case, the $A_{II}$-$B_{II}$ pair exhibits non-monotonic behavior with a peak whose position shifts to larger $a$ as $\omega$ increases, and both parameters contribute only weakly except near an optimal $\frac{\omega}{s}$ ratio.
\end{abstract}

\maketitle

\section{Introduction}
What quantity can fully characterize the quantum correlations in a given system? The conventional answer is entanglement. However, recent studies have shown that separable (i.e., non-entangled) states, traditionally referred to as classically correlated states, can still retain certain quantum features with potential applications in quantum technologies \cite{Datta2008PRL,Lanyon2008PRL,Piani2008PRL,Ferraro2010PRA,Brodutch2010PRA}. One such feature is quantum discord \cite{ollivier2001quantum}.
Quantum discord is a physical quantity that characterizes quantum correlations based on measurements. It holds a fundamental position in quantum information science and has a very wide range of applications \cite{luo2008using,li2008classical,radhakrishnan2020multipartite,inui2020entanglement,zhou2020quantum}.
Unlike entanglement, quantum discord can describe a broader class of nonclassical correlations: it takes a nonzero value even for separable states, and generally exhibits stronger robustness against decoherence than entanglement.
Analytical solutions for this measure have been derived for families of two-qubit states with special structures \cite{geometric_measure_quantum_discord,quantum_discord_two_qubit_systems,quantum_discord_geometry_bell_diagonal,quantum_discord_geometry_class_two_qubit,quantum_discord_two_qubit_analytical_small_error,quantum_discord_x_states_one_variable,analytical_formula_quantum_discord_two_qubit_x,super_quantum_discord_two_qubit_x,computing_quantum_discord_np_complete}.
For multipartite quantum systems, researchers have proposed generalized forms including the squared quantum discord and multiqubit geometric discord \cite{exploring_multipartite_quantum_correlations_square_discord,quantum_discord_multiqubit_systems,geometric_discord_multiqubit_systems,monogamy_deficit_quantum_correlations_multipartite,conditions_monogamy_quantum_correlations_ghz_w}.
For continuous-variable states, Adesso \textit{et al.} derived a general analytical expression for Gaussian discord \cite{gaositaizhongdeliangziyujingdianguanlian}, while Giorda \textit{et al.} obtained a closed-form expression for squeezed thermal states \cite{giorda_paris_2010}.

Recent work in relativistic quantum information has focused on non-inertial spacetimes, with extensive studies investigating the behavior of quantum correlations, including entanglement \cite{zhang2025entanglement,liu2025entanglement,liu2023fermionic,wang2020genuine,mi2024impact,Li2023Quantum,feiguanxingxizhongdelianxubianliangjiuchangongxiang,limanshikongxiahuojin-anluxiaoyingzuoyongyushuangmoyasutajiuchantaideyanjiu}, coherence \cite{li2025multiqubit,wu2021quantum,liao2025quantum,Kaczmarek2025Coherence}, quantum correlations \cite{liangziyujingdianguanliandeturanbianhuayijianluxiaoying,xiangduilunliangzixinxizhongdelianxubianliangfangfa}, nonlocality \cite{mi2025genuine,zhang2023hawking,Kaczmarek2024Signatures,Kaczmarek2026Nonlocal}, uncertainty relations \cite{Wang2024Entropic},  quantum steering \cite{wu2025fermionic,wu2025gaussian,liu2018influence,Mi2026Gaussian}, quantum discord \cite{huang2026,non_markovian_effect_quantum_discord,sudden_change_quantum_discord_single_qubit_noise,quantum_discord_resource_remote_state_preparation,observing_operational_significance_discord_consumption,quantum_discord_bounds_distributed_entanglement,quantum_coherence_geometric_quantum_discord,converting_coherence_quantum_correlations,quantum_coherence_multipartite_systems,unified_view_quantum_correlations_quantum_coherence,relative_quantum_coherence_incompatibility_quantum_correlations_states} and so on. These studies deepen our understanding of quantum information in non-inertial backgrounds and further advance research on the black hole information paradox and entanglement entropy.
Despite considerable progress in this area, the dynamical evolution of continuous-variable quantum discord under the Unruh effect remains insufficiently understood.

In this paper, we study the redistribution of Gaussian quantum discord under the Unruh effect. We consider the discord between two field modes, each described from a different observer's perspective, with Alice and Bob initially sharing a two-mode squeezed Gaussian state. The Unruh effect transforms the vacuum state of a non-inertial observer into a thermal state \cite{Davies1975JPhysA,Unruh1976PRD}. From the joint perspective of one inertial and one uniformly accelerated observer, the system is described by a three-mode state; when both observers accelerate, it becomes a four-mode state. A physical observer undergoing uniform acceleration can access only one of the two non-inertial modes. As a result, tracing over the inaccessible modes to obtain the reduced state reveals that some correlations are lost. Here we consider a free scalar field prepared in a two-mode squeezed state as seen from the inertial frame. This choice is motivated by its relevance to a range of physical settings. First, the two-mode squeezed Gaussian state is chosen for two reasons: it is a canonical continuous-variable entangled state approximating EPR pairs with arbitrary precision, and it is experimentally accessible for existing continuous-variable quantum information setups \cite{Einstein1935EPR,Braunstein2005RMP}. Second, the Unruh-induced transformation admits an equivalent description as a Gaussian channel acting on the initial Gaussian state within the quantum information framework.

The rest of this paper is organized as follows. Section II is devoted to a review of the Gaussian channel picture of the Unruh effect and the formal framework of continuous-variable quantum discord. In Section III, we systematically investigate the dynamical characteristics of quantum discord under the Unruh effect and present the main findings on its evolutionary behavior. At last, Section IV gives a brief summary of the whole work.

\section{The Unruh effect via Gaussian channels and Gaussian quantum discord}

\subsection{Unruh effect by Gaussian channels}

First, we consider an observer moving with uniform acceleration $a$ in the $(t,z)$ plane 
(with the unit convention $c=1$). Rindler coordinates $(\tau,\zeta)$ are well suited 
to describe the perspective of a uniformly accelerated observer. Two distinct sets 
of Rindler coordinates, differing by an overall sign, are necessary to cover 
the full Minkowski spacetime.
\begin{align}
	a t &= e^{a\zeta} \sinh(a\tau),\quad a z = e^{a\zeta} \cosh(a\tau), \\
	a t &= -e^{a\zeta} \sinh(a\tau),\quad a z = -e^{a\zeta} \cosh(a\tau).	
	\label{eq:1-2}
\end{align}

These two coordinate patches define two causally disconnected Rindler regions, 
conventionally denoted $I$ and $II$. A particle undergoing eternal uniform 
acceleration is confined to either region $I$ or $II$, with no causal access 
to the complementary region.

We now consider a free quantum scalar field on a flat spacetime background. 
Quantization of the scalar field in Minkowski coordinates is not equivalent 
to that in Rindler coordinates. Nevertheless, from the Rindler perspective 
\cite{Davies1975JPhysA,Unruh1976PRD}, the vacuum state of a given field mode as described 
by an inertial observer can be cast as a two-mode squeezed state \cite{WallsMilburn1994,Fulling1973PRD}.
\begin{equation}
	\begin{split}
		\ket{0_k}_{M}
		&= \frac{1}{\cosh r_k} \sum_{n=0}^{\infty} \tanh^n r_k \, \ket{n_k}_{I}\ket{n_k}_{{II}} \\
		&= U_k \ket{0_k}_{I}\ket{0_k}_{{II}}.
		\label{eq3}
	\end{split}
\end{equation}
Here, 
\begin{equation}
	U_k = \exp\left[ r_k \left( \hat{b}^\dagger_{k,I} \hat{b}^\dagger_{k,II} - \hat{b}_{k,I} \hat{b}_{k,II} \right) \right]
	\label{uk}
\end{equation}
stands for a two-mode squeezing operator, where the squeezing parameter $r_k$ is fixed by the algebraic relation
\begin{equation}
	\cosh r_k = \left( 1 - e^{-\frac{2\pi |\omega_k|}{a}} \right)^{-\frac{1}{2}},
	\label{eqr}
\end{equation}
 in which $a$ is the uniform acceleration and $\omega_k$ is the mode frequency. $|n_k\rangle_{I}$ and $|n_k\rangle_{{II}}$ are the mode decompositions in Rindler regions $I$ and $II$, respectively. To streamline subsequent derivations, we introduce
\begin{equation}
	\theta_k^2 =\tanh^2 r_k =e^{ -\frac{2\pi |\omega_k|}{a}}.
	\label{eq4}
\end{equation}

Each Minkowski mode with frequency $|\omega_k|$ admits a Rindler mode expansion as given by Eq.~\eqref{eq3}.
The relations for higher energy states can be derived from Eq.~\eqref{eq3}
together with the Bogoliubov transformation between the creation and annihilation operators.
\begin{equation}
	\begin{split}
	\hat{a}_k &= \cosh r_k \, \hat{b}_{I,k} - \sinh r_k \, \hat{b}^\dagger_{{II,k}}, \\
	\hat{a}_k^\dagger &= \cosh r_k \, \hat{b}_{I,k}^\dagger  -   \sinh r_k \, \hat{b}_{II,k},	
	\end{split}
	\label{eq5}
\end{equation}
where $\hat{a}_k$ and $\hat{a}_k^\dagger$ denote the annihilation and creation operators for mode $k$ in Minkowski space, respectively. For the same field mode, $\hat{b}_{I,k}$ and $\hat{b}_{II,k}$ are the annihilation operators in Rindler regions $I$ and $II$ \cite{Davies1975JPhysA,Unruh1976PRD}, while $\hat{b}_{I,k}^\dagger$ and $\hat{b}_{II,k}^\dagger$ are the corresponding creation operators.
A Rindler observer moving in region $I$ needs to trace over the modes in region $II$,
since the observer has no access to the information in this causally disconnected region.
Therefore, while a Minkowski observer concludes that the field mode $k$
is in the vacuum state $|0_k\rangle_{M}$, the state from the perspective
of an observer in uniform acceleration $a$, constrained to region $I$, is
\begin{equation}
\ket{0_k}\bra{0_k}_{M} \to \frac{1}{\cosh^2 r_k} \sum_{n=0}^{\infty} \tanh^{2n} r_k \, \ket{n_k}\bra{n_k}_{I},
\label{eq6}
\end{equation}
which is a thermal state with temperature $T = \frac{a}{2\pi k_\mathrm{B}}$, where $k_\mathrm{B}$ is Boltzmann's constant.

The squeezing operator $U_k$ is a Gaussian operation that preserves the Gaussianity of input states.
Therefore, Eq.~\eqref{eq3}
demonstrates that the vacuum state transformation underlying the Unruh effect
can be described by a Gaussian channel.
In phase space, the two-mode squeezing operator $U_k$ corresponds to a symplectic transformation
\begin{equation}
	S_k = \frac{1}{\sqrt{1-\theta_k^2}}
	\begin{pmatrix}
		I_2 & \theta_k Z_2 \\
		\theta_k Z_2 & I_2
	\end{pmatrix},
	\label{eq:symplectic_squeezing}
\end{equation}
where $I_2$ denotes the unity matrix in $2\times2$ space, and $Z_2  =\begin{pmatrix}
	1 & 0\\
	0 & -1
\end{pmatrix}.$

\subsection{Quantum discord of Gaussian states}

Within the present subsection, we survey core definitions and symbolic conventions relevant to Gaussian states. We focus on an $n$-mode continuous-variable system equipped with state space $\mathcal{H}=\mathcal{H}_1\otimes \mathcal{H}_2 \otimes \dots \otimes \mathcal{H}_n$, where each subspace $\mathcal{H}_i$ satisfying $1 \leq i \leq n$ corresponds to an infinite-dimensional complex Hilbert space. For each $m=1,2,\dots,n$, we label the single-mode Fock basis as the set $\{|j_m\rangle\}_{j_m=0}^{\infty} \subset \mathcal{H}_m$.

Denote by $\mathcal{S(H)}$  the set of all quantum states (that is, positive bounded linear operators with trace 1) on $\mathcal{H}$.
For any state $\rho\in \mathcal{S(H)}$, its characteristic function $\chi_{\rho}$ can be expressed as
\begin{equation}
	\chi_{\rho}(z)=\rm{Tr} \rho W(z),
\end{equation}
where $z=(x_1, y_1, x_2, y_2, \dots, x_n, y_n)^\mathrm{T}$, $R=(\hat{R}_1, \hat{R}_2, \dots, \hat{R}_{2n}) \equiv (\hat{X}_1, \hat{Y}_1, \dots, \hat{X}_n, \hat{Y}_n)$, and $W(z)=\exp(\mathrm{i} R^\mathrm{T} z)$ denotes the Weyl displacement operator. The superscript $\mathrm{T}$ stands for transposition. Here, as usual, $\hat{X}_p=\hat{c}_p+\hat{c}_p^{\dagger}$, and  
$\hat{Y}_p=\mathrm{i}(\hat{c}_p-\hat{c}_p^{\dagger})$ ($p=1, 2, \dots, n$) respectively stand for the position and momentum operators,
with  $\hat{c}_p^{\dagger}$ and 
$\hat{c}_p$  the creation and annihilation
operators in the $p$-th  mode $\mathcal{H}_p$ satisfying the canonical commutation relation
\[
[\hat{c}_p, \hat{c}_q^{\dagger}]=\delta_{pq}I, ~~[\hat{c}_p^{\dagger}, \hat{c}_q^{\dagger}]=[\hat{c}_p, \hat{c}_q]=0,
\]
where $ p, q=1, 2, \dots, n$.

\begin{figure}
	\centering
	\includegraphics[width=0.8\linewidth]{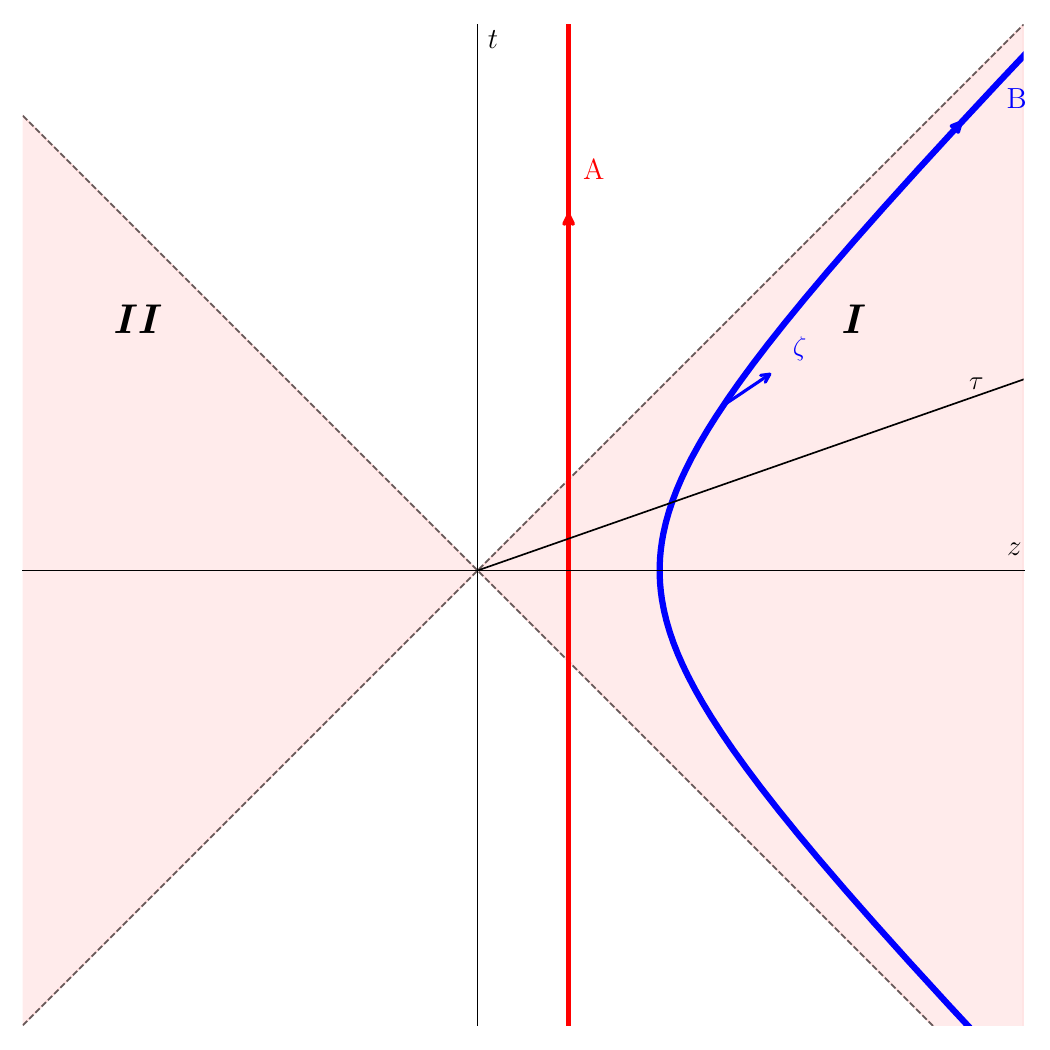}
	\caption{Schematic diagram of the worldlines for the inertial observer Alice and the uniformly accelerated observer Bob. The set $(z,t)$ denotes the Minkowski coordinates, while the set $(\zeta,\tau)$ denotes the Rindler coordinates. The causally disconnected Rindler regions $I$ and $II$ are indicated.}
	\label{fig:Bjiasu}
\end{figure}

Assume that $\rho$ has finite second moment. The displacement vector (or mean) $\bar{d}$ of $\rho$  is given by
\begin{equation}
	\bar{d}=(\langle \hat{R}_1 \rangle, \dots, \langle \hat{R}_{2n}\rangle)
	=(\mathrm{Tr}(\rho \hat{R}_1), \dots, \mathrm{Tr}(\rho \hat{R}_{2n}))     
\end{equation}
and the covariance matrix $\sigma=(\sigma_{pq})$ of $\rho $ is defined as
\begin{equation}
	\sigma_{pq}=\frac{1}{2}\langle  \Delta \hat{R}_p \Delta \hat{R}_q+\Delta \hat{R}_q\Delta \hat{R}_p\rangle,   
\end{equation}
where $\Delta \hat{R}_p=\hat{R}_p-\langle \hat{R}_p \rangle$ \cite{Braunstein2005QICV}. 
It should be noted that a covariance matrix $\sigma$ is symmetric and must satisfy the uncertainty principle
\[
\sigma+\mathrm{i}\Omega_n\geq 0,
\]
where $\Omega_n=\oplus_{i=1}^n\Omega_i$ with 
$\Omega_i=\begin{pmatrix} 0 & 1\\-1 & 0    \end{pmatrix}$ for each $i$ \cite{Simon1994QNM}. 
In addition, $\sigma \ge 0$ as $\sigma +\mathrm{i}\Omega_n \ge 0$.

Every property possessed by a Gaussian state is uniquely fixed by the first and second statistical moments associated with the quadrature operators. Local unitary operations enable arbitrary tuning of the first moments without altering any informationally meaningful features, such as entropy and various correlation quantifiers. Given this invariance property, one may safely fix all first moments to zero; accordingly, the second-order moments constitute the only necessary descriptors for Gaussian states.

Based on our research, we initially consider a two-mode Gaussian state $\rho$ shared by Alice and Bob, which is 
specified by its covaricance matrix, a real,  symmetric and 
positive matrix:
\begin{equation}
	\sigma = \begin{pmatrix}
		\mathcal{A} & \mathcal{C} \\
		\mathcal{C}^\mathrm{T} & \mathcal{B}
	\end{pmatrix},
	\label{eq:cov_sts}
\end{equation}
where $\mathcal{A} = \mathrm{diag}(a,a)$, $\mathcal{B} = \mathrm{diag}(b,b)$, and $\mathcal{C} = \mathrm{diag}(c_{1},c_{2})$ are $2\times2$ diagonal matrices.

The quantum discord of a Gaussian state can be expressed as \cite{gaositaizhongdeliangziyujingdianguanlian}
\begin{equation}
	D(\sigma) = h\left(\sqrt{\lambda_2}\right) - h(d_-) - h(d_+) + \inf_{\sigma_0} h\left(\sqrt{\det{\sigma_P}}\right).
	\label{eq:gaussian_discordyiban}
\end{equation}
\begin{figure}
	\centering
	\includegraphics[width=0.8\linewidth]{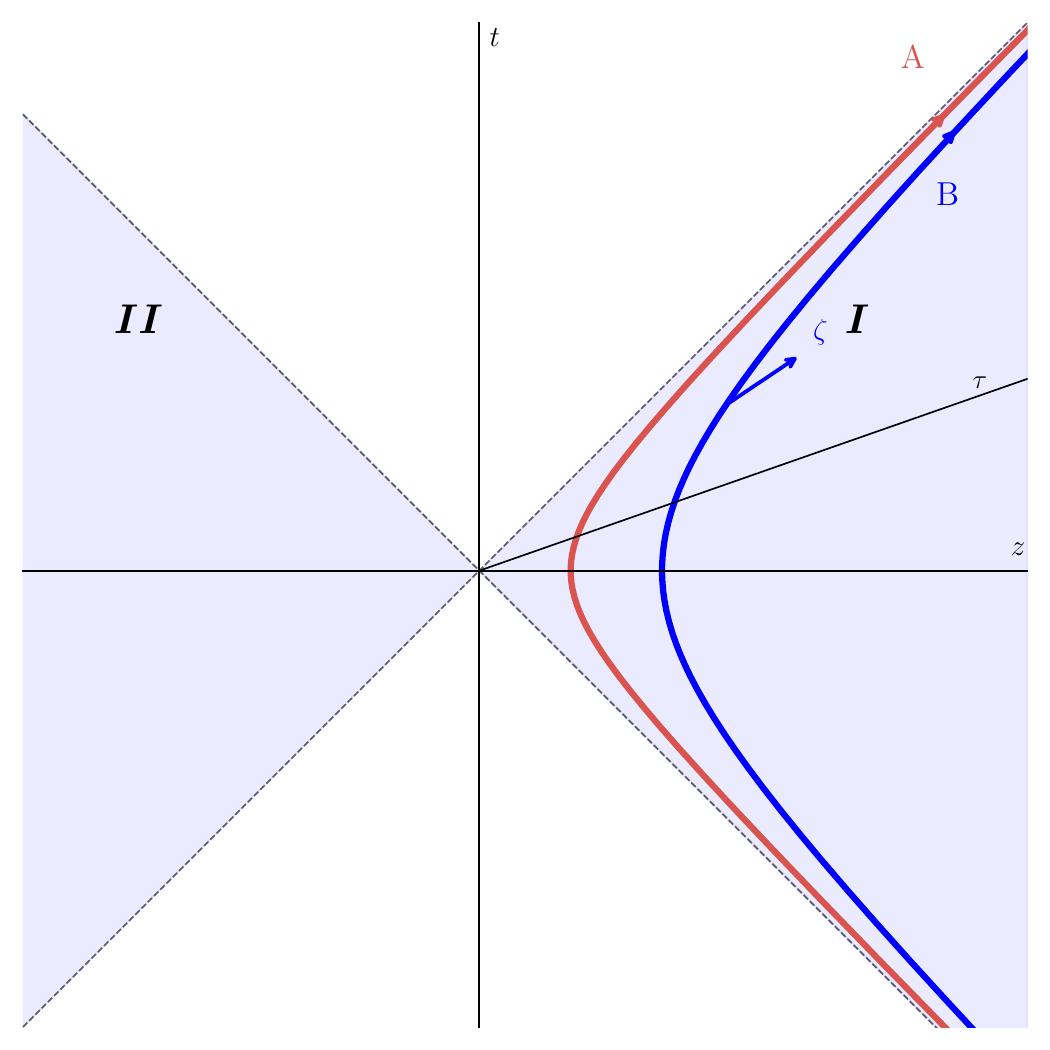}
	\caption{Schematic diagram of the worldlines for two uniformly accelerated observers, Alice and Bob. The set $(z,t)$ denotes the Minkowski coordinates, while the set $(\zeta,\tau)$ denotes the Rindler coordinates. The causally disconnected Rindler regions $I$ and $II$ are indicated.}
	\label{fig:ABjiasu}
\end{figure}
Here, $
\sigma_P = \mathcal{A} - \mathcal{C} \left( \mathcal{B} + \sigma_0 \right)^{-1} \mathcal{C}^\mathrm{T},
\label{eq:schur_complement}
$
where $
\sigma_0 = 
\begin{pmatrix}
	\alpha & \gamma \\
	\gamma & \beta
\end{pmatrix},
$
with fixed parameters $\alpha, \beta \in \mathbb{R}^+$ and $\gamma \in \mathbb{R}$.
 The  function takes the form
\begin{equation}
	h(x) = \left(x+\frac{1}{2}\right)\ln\left(x+\frac{1}{2}\right) - \left(x-\frac{1}{2}\right)\ln\left(x-\frac{1}{2}\right),
	\label{eq:h_function}
\end{equation}
as well as the quantities $d_\pm$ satisfying
\begin{equation}
	d_\pm^2 = \frac{1}{2}\left( \Delta \pm \sqrt{\Delta^2 - 4\lambda_4} \right),
	\Delta = \lambda_1 + \lambda_2 + 2\lambda_3,
	\label{eq:d_pm_def}
\end{equation}
where $\lambda_1 = \det\mathcal{A}$, $\lambda_2 = \det\mathcal{B}$,  $\lambda_3 = \det\mathcal{C}$, and  $\lambda_4 = \det\sigma$.

For the generic bipartite squeezed thermal states (STS), the matrix elements of the covariance matrix \eqref{eq:cov_sts} are given by
\begin{equation}
	\begin{split}
		&a = \frac{1}{2} + N_\xi + N_1(1+N_\xi) + N_2 N_\xi, \\
		&b = \frac{1}{2} + N_\xi + N_2(1+N_\xi) + N_1 N_\xi, \\
		&c_1 = -c_2 = \left(1 + N_1 + N_2\right)\sqrt{N_\xi(1+N_\xi)},
	\end{split}\notag
	\label{eq:cov_elements_sts}
\end{equation}
where $N_\xi=\sinh^2 \xi$ and $N_j$ is the mean 
thermal photon number of mode $j$, and $j\in \{1, 2\}$.

The quantum discord of the generic bipartite squeezed thermal states can be expressed as \cite{giorda_paris_2010}
\begin{equation}
	\begin{split}
		D(\sigma) &= h(\sqrt{\lambda_2}) - h(d_-) - h(d_+) \\
		&+ h\left( \frac{\sqrt{\lambda_1} + 2\sqrt{\lambda_1 \lambda_2} + 2\lambda_3}{1 + 2\sqrt{\lambda_2}} \right).
	\end{split}
	\label{eq:quantum_discord_sts}
\end{equation}

\section{Influence of the Unruh Effect on Gaussian Quantum Discord}
\begin{figure*}
	\centering
	\subfloat[$A$-$B_I$]{\includegraphics[width=0.32\textwidth]{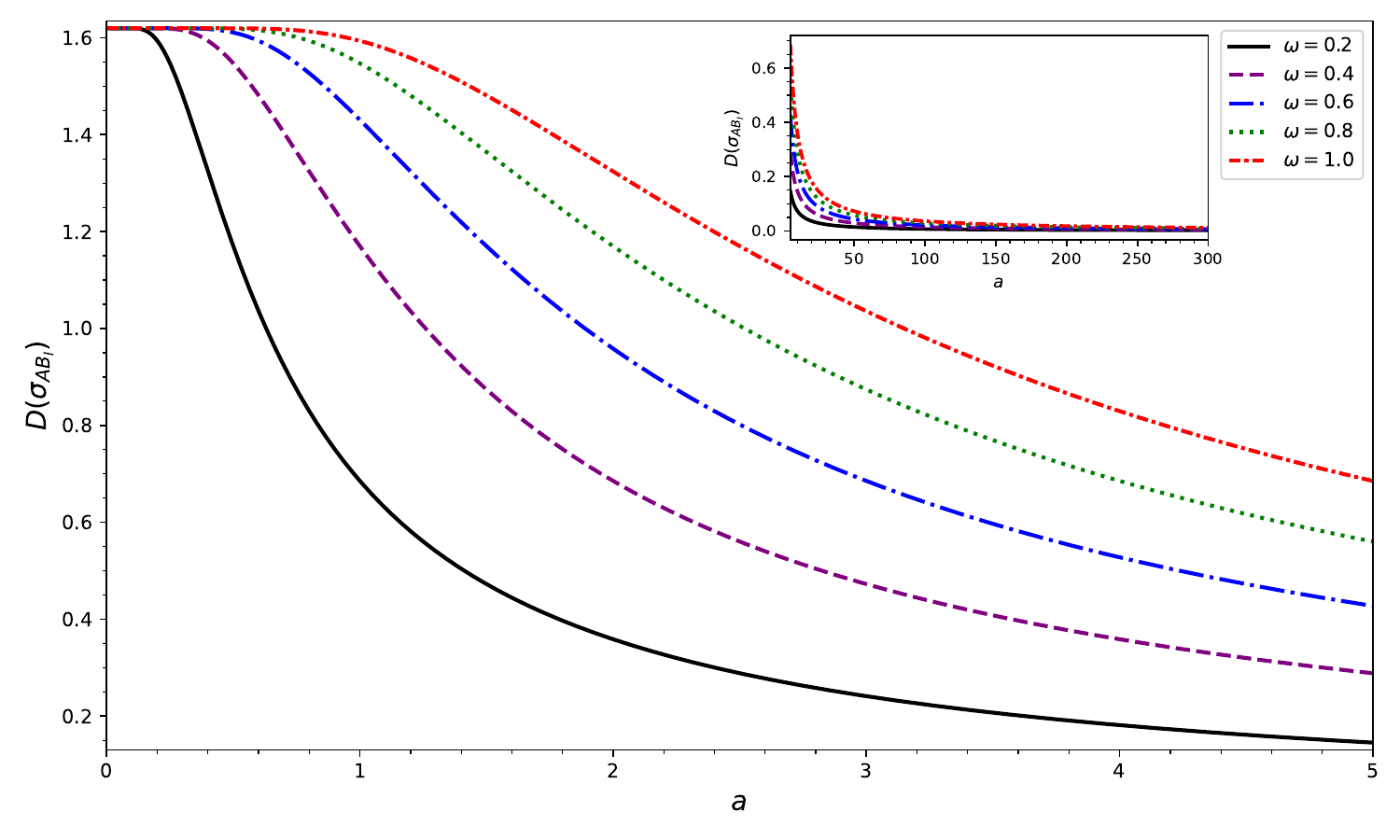}}
	\hfill
	\subfloat[$B_I$-$B_{II}$]{\includegraphics[width=0.32\textwidth]{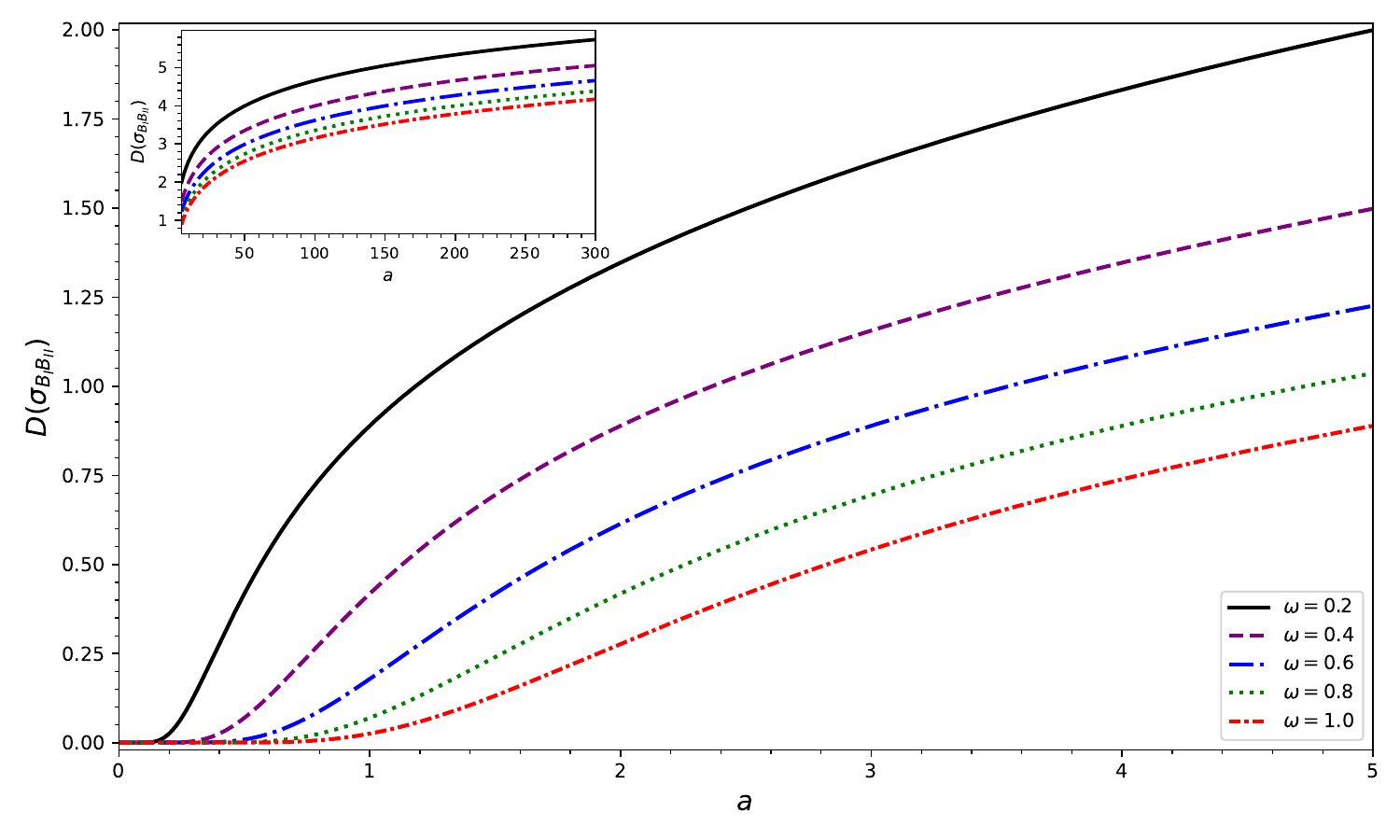}}
	\hfill
	\subfloat[$A$-$B_{II}$]{\includegraphics[width=0.32\textwidth]{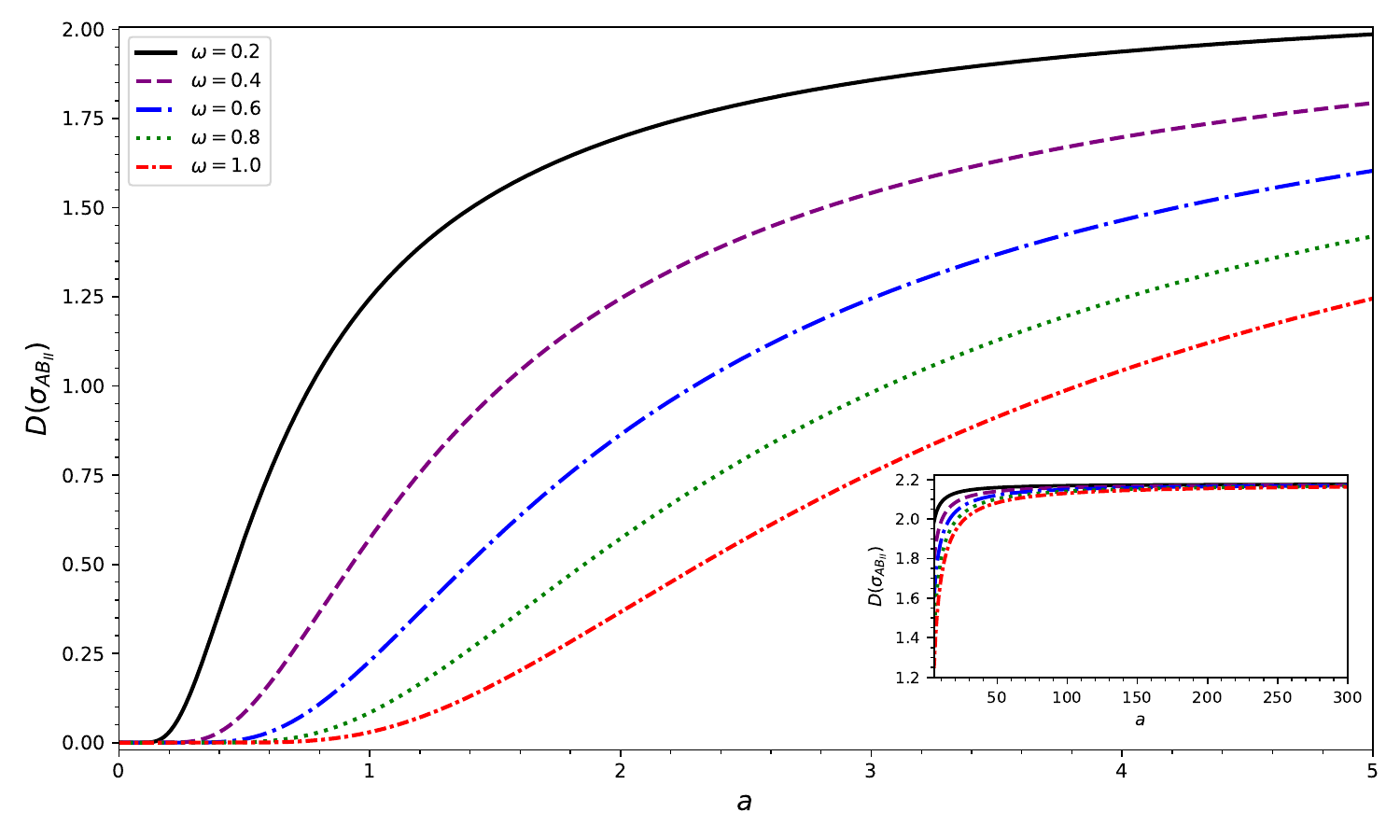}}
	
	\vspace{0em} 
	
	\subfloat[$A$-$B_I$]{\includegraphics[width=0.32\textwidth]{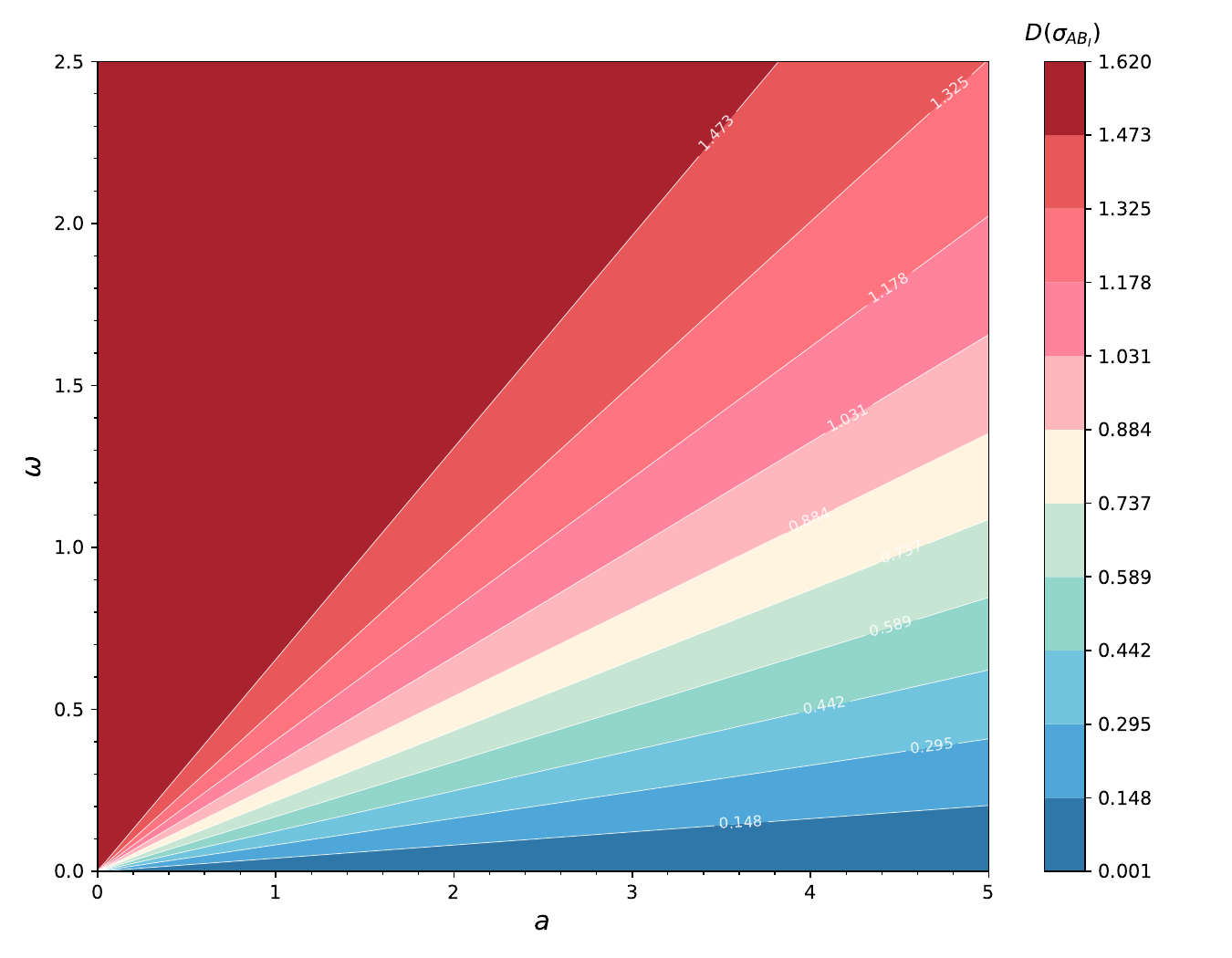}}
	\hfill
	\subfloat[$B_I$-$B_{II}$]{\includegraphics[width=0.32\textwidth]{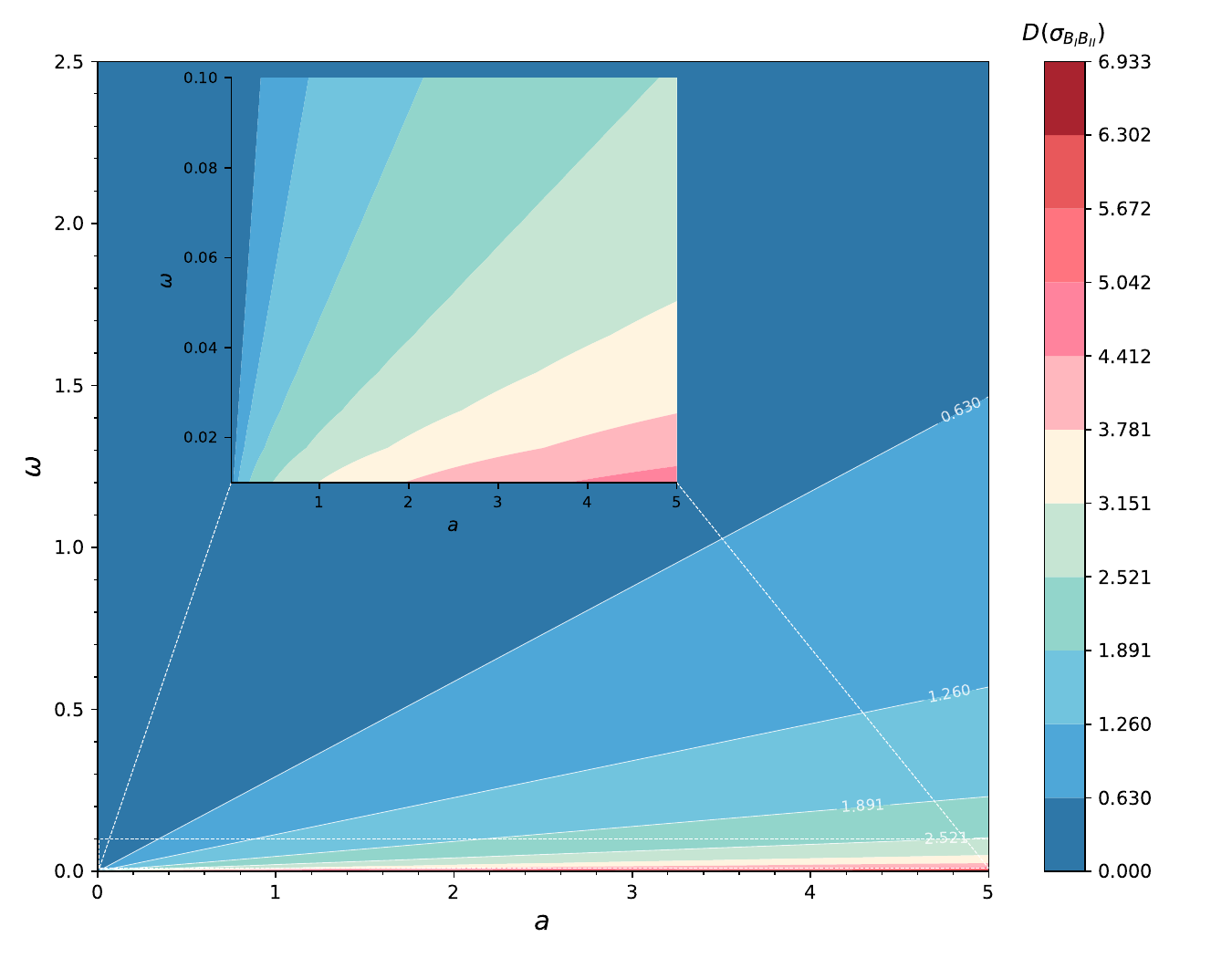}}
	\hfill
	\subfloat[$A$-$B_{II}$]{\includegraphics[width=0.32\textwidth]{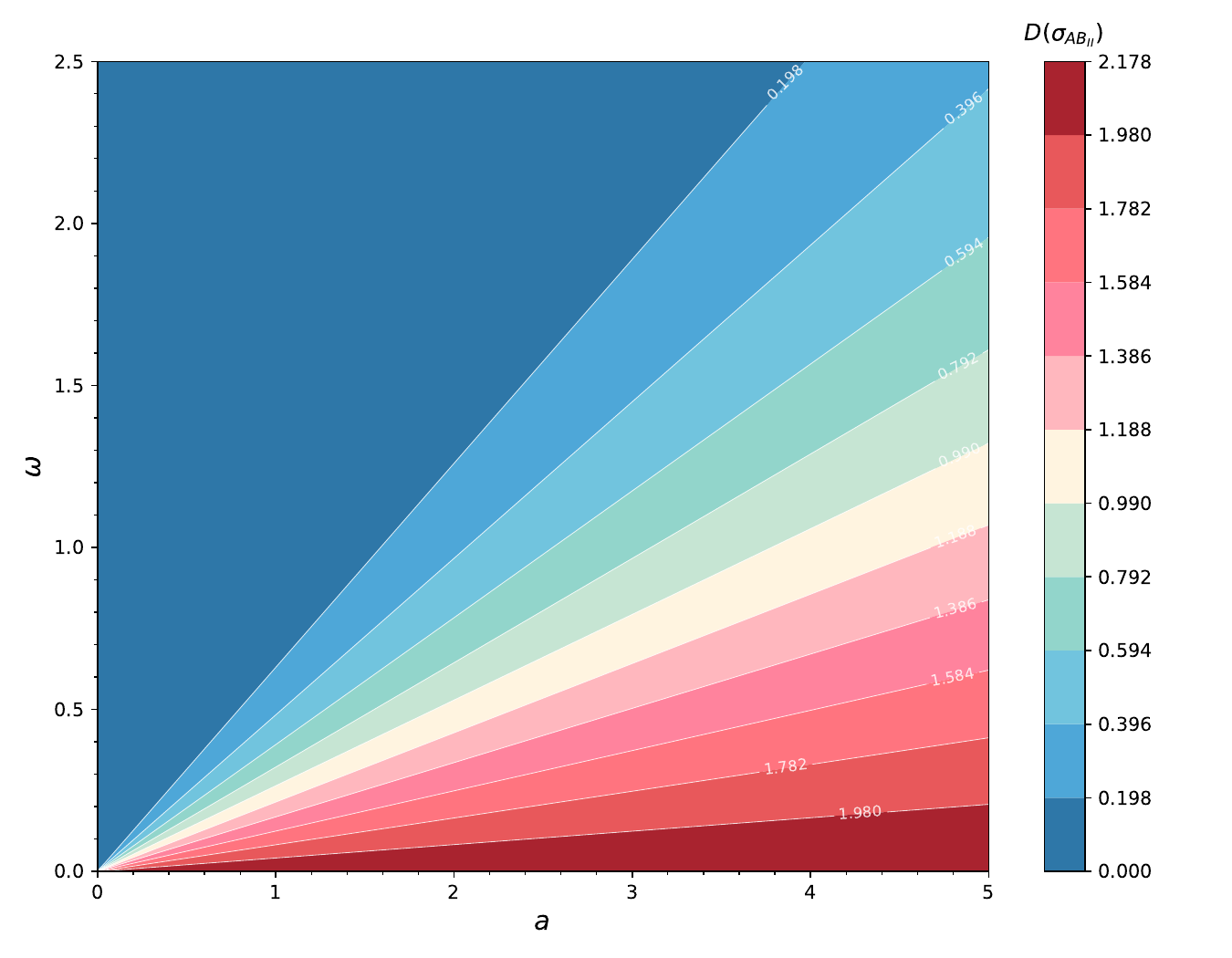}}
	
	\caption{Gaussian quantum discord for different mode pairs as a function of the acceleration $a$ and the field frequency $\omega$, with the squeezing parameter fixed at $s=1$.}
	\label{fig:2D-ABI-BIBII-ABII}
\end{figure*}
In this section, we consider the scenario where Alice and Bob initially share a pure two-mode squeezed Gaussian state. The corresponding covariance matrix reads \cite{WallsMilburn1994,adesso_fuentes_schuller_ericsson_2007}:
\begin{equation}
	\sigma_{AB} =
	\begin{pmatrix}
		\dfrac{\cosh(2s)}{2}\, I_2 & \dfrac{\sinh(2s)}{2}\, Z_2 \\[1.2em]
		\dfrac{\sinh(2s)}{2}\, Z_2 & \dfrac{\cosh(2s)}{2}\, I_2
	\end{pmatrix},
	\label{eq:tmss_covariance}
\end{equation}
where $s$ denotes the squeezing parameter.

\subsection{Gaussian quantum discord distribution induced by a uniformly accelerated observer}

\begin{figure}
	\centering
	\begin{minipage}[b]{0.9\columnwidth}
		\centering
		\includegraphics[width=\linewidth,trim={10 0 0 100},clip]{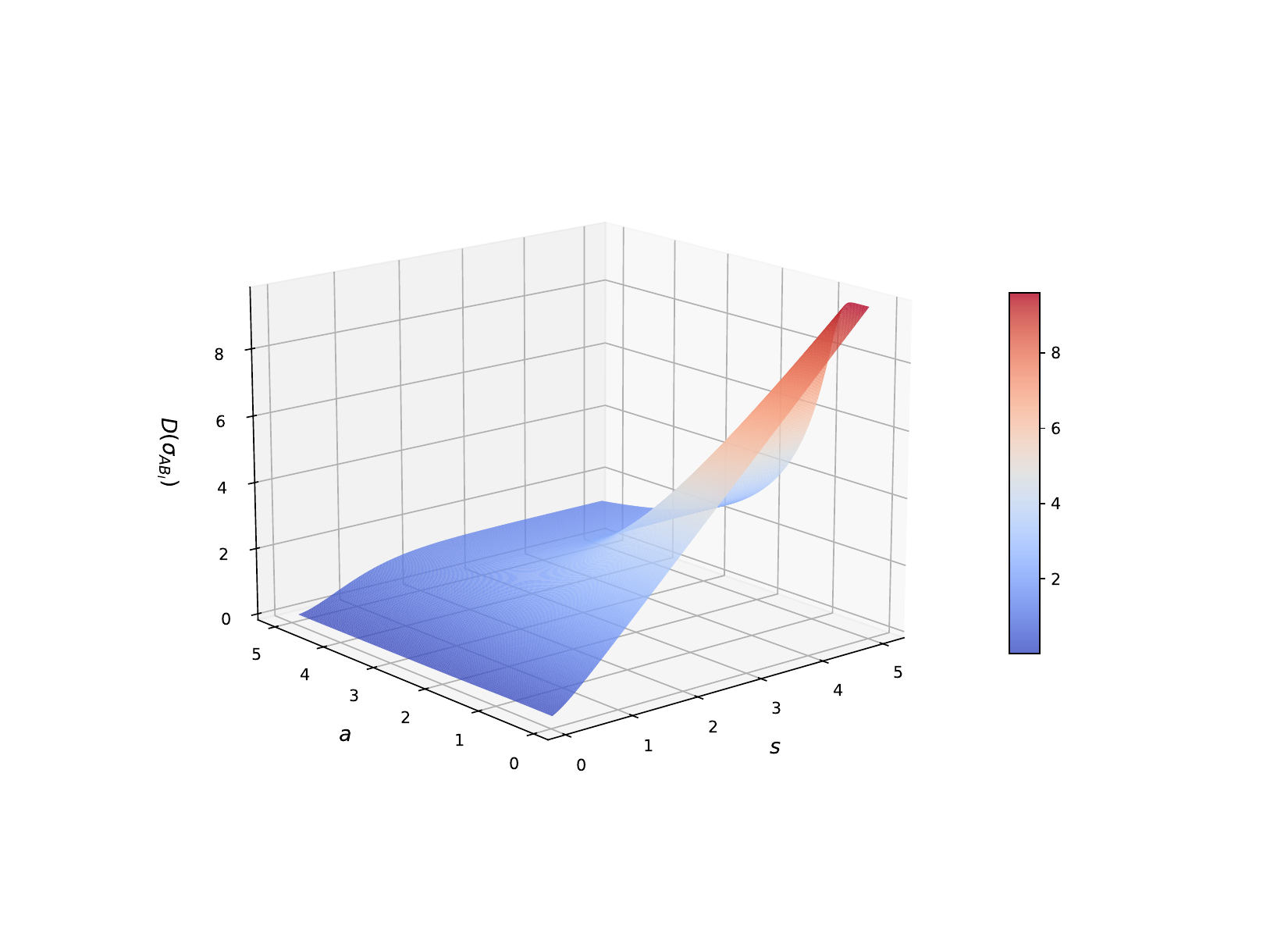}
		\\[-3.5em]  
		\hspace*{-2.7em}(a)
	\end{minipage}
	\hfill
	\begin{minipage}[b]{0.7\columnwidth}
		\centering
		\includegraphics[width=\linewidth]{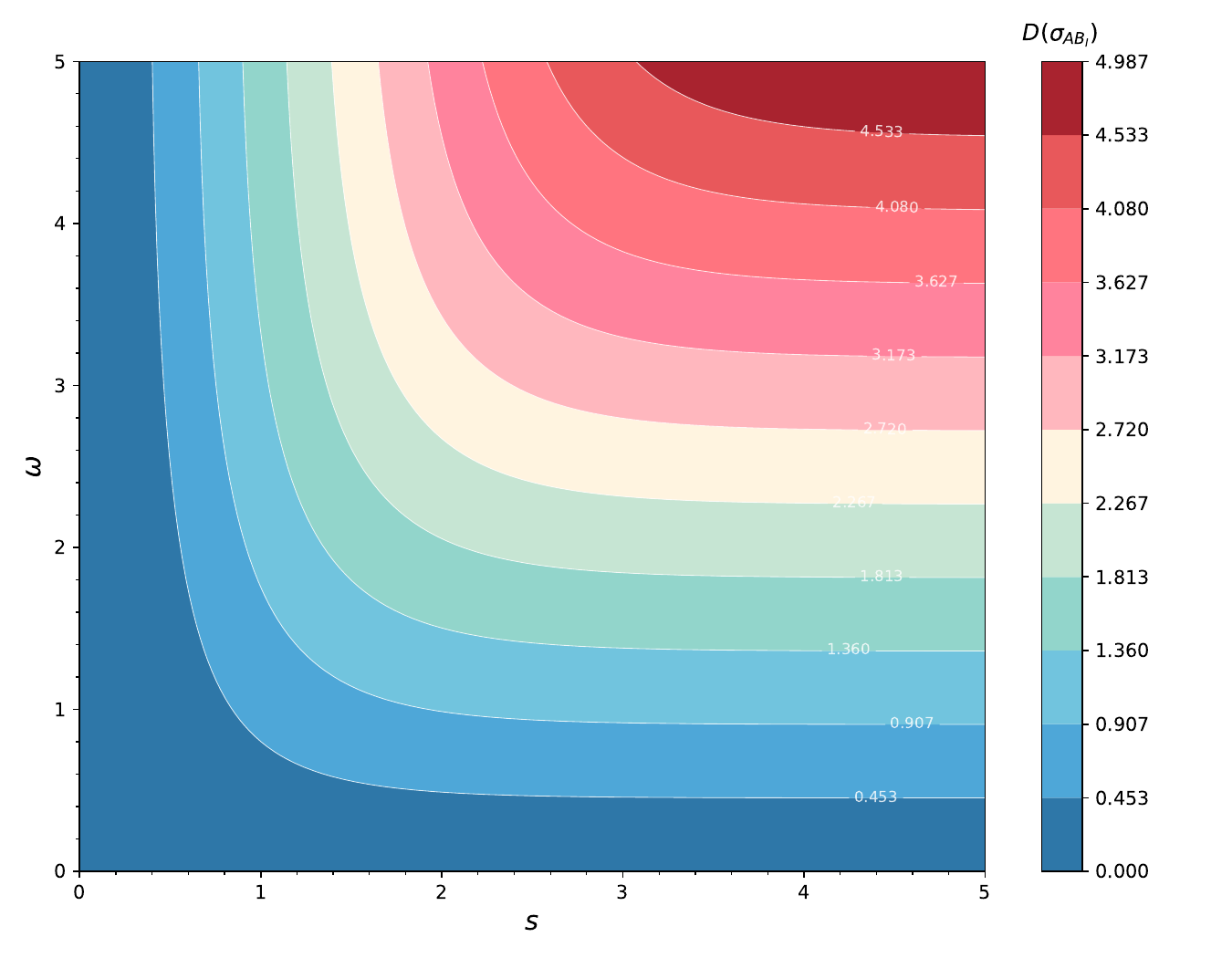}
		\\[-0.5em]
		\hspace*{-2.7em}(b)
	\end{minipage}
	\caption{Gaussian quantum discord of the mode pair $A$-$B_{I}$. (a) Discord as a function of the acceleration $a$ and the squeezing parameter $s$ for a fixed field frequency $\omega=1$. (b) Discord as a function of the field frequency $\omega$ and the squeezing parameter $s$ for a fixed acceleration $a=2\pi$.}
	\label{fig:3D-ABI}
\end{figure}

\begin{figure}
	\centering
	\begin{minipage}[b]{0.9\columnwidth}
		\centering
		\includegraphics[width=\linewidth,trim={10 0 0 100},clip]{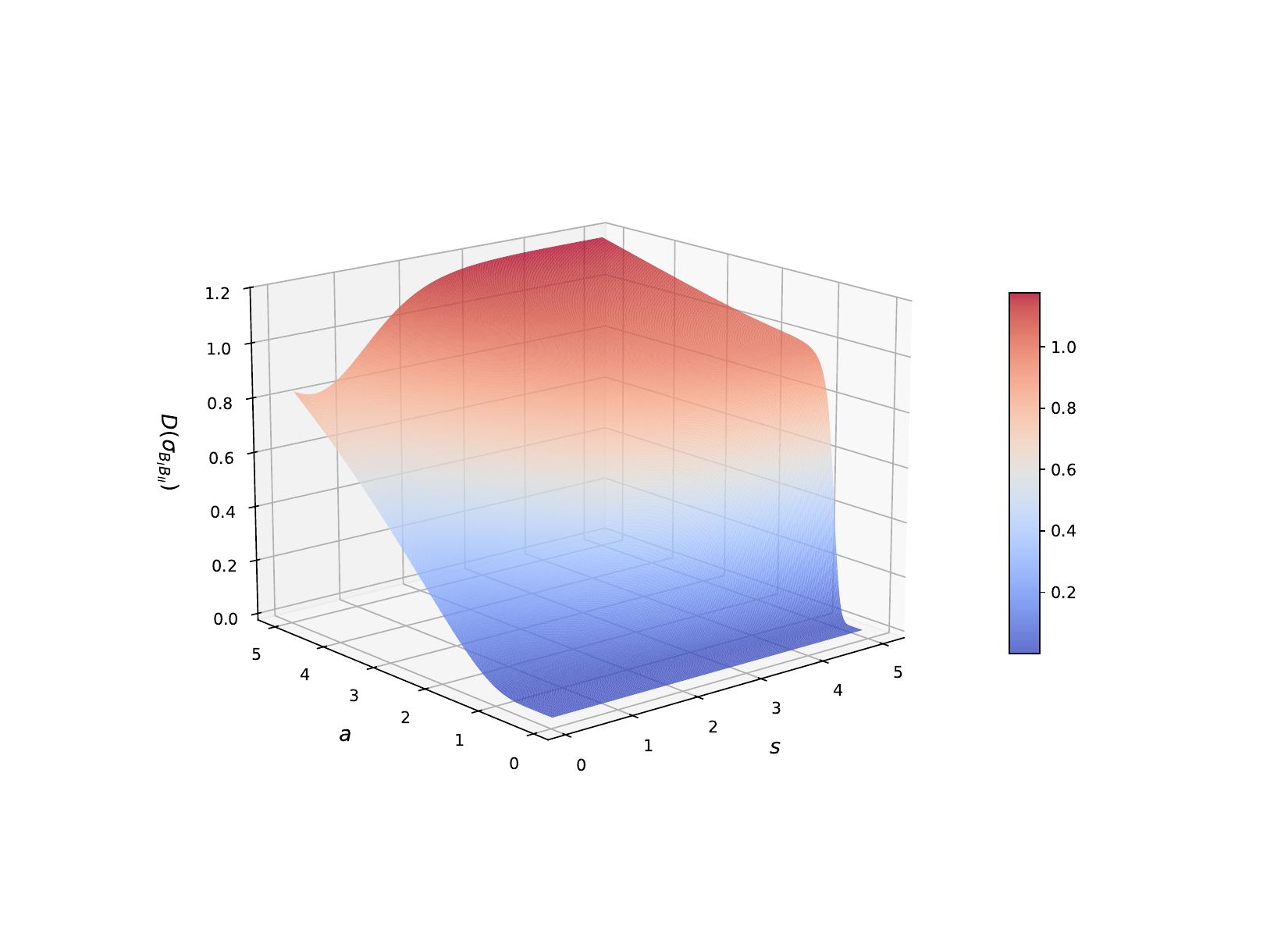}
		\\[-3.5em]  
		\hspace*{-2.7em}(a)
	\end{minipage}
	\hfill
	\begin{minipage}[b]{0.7\columnwidth}
		\centering
		\includegraphics[width=\linewidth]{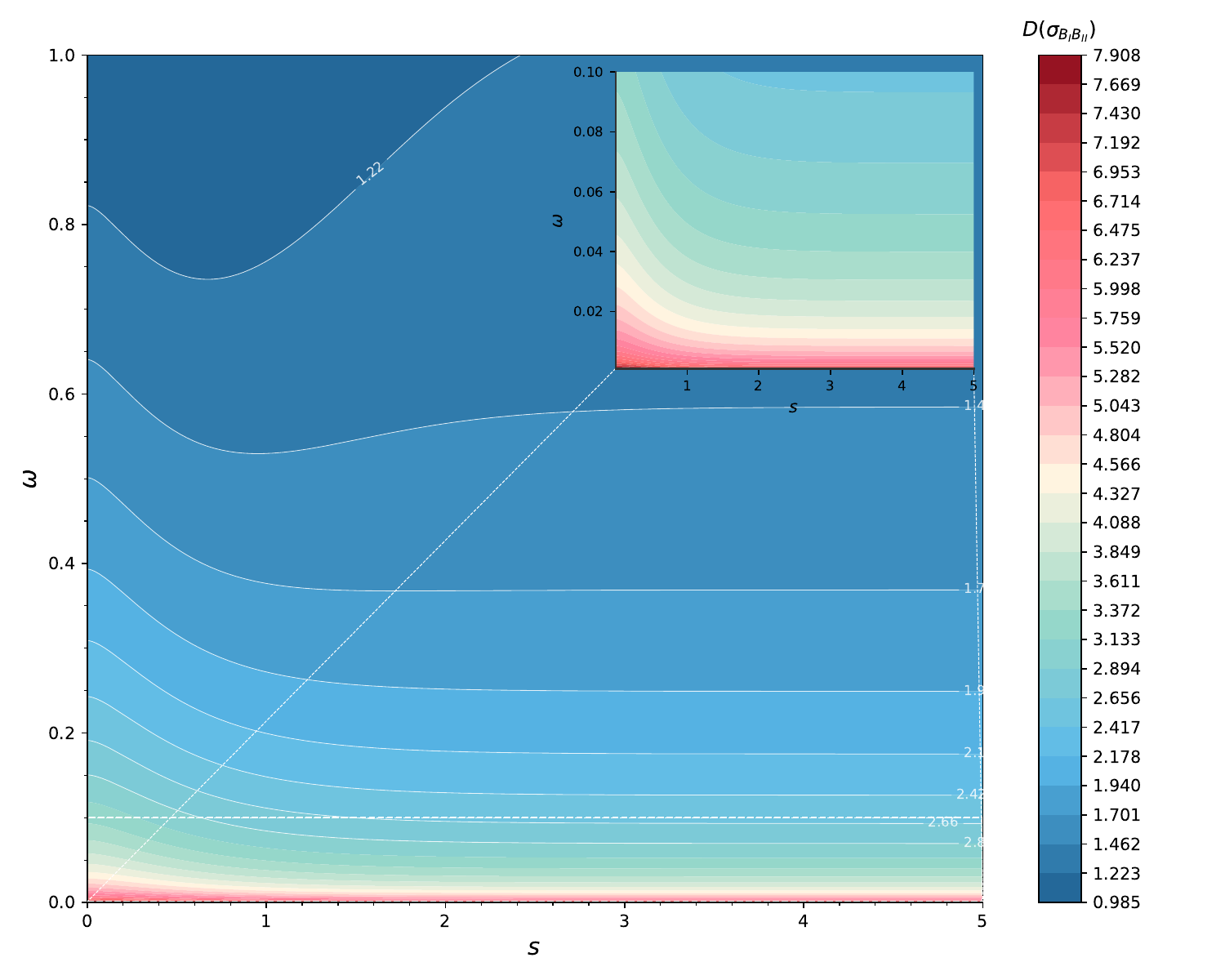}
		\\[-0.5em]
		\hspace*{-2.7em}(b)
	\end{minipage}
	\caption{Gaussian quantum discord of the mode pair $B_I$-$B_{II}$. (a) Discord as a function of the acceleration $a$ and the squeezing parameter $s$ for a fixed field frequency $\omega=1$. (b) Discord as a function of the field frequency $\omega$ and the squeezing parameter $s$ for a fixed acceleration $a=2\pi$.}
	\label{fig:3D-BIBII}
\end{figure}

\begin{figure}
	\centering
	\begin{minipage}[b]{0.9\columnwidth}
		\centering
		\includegraphics[width=\linewidth,trim={10 0 0 100},clip]{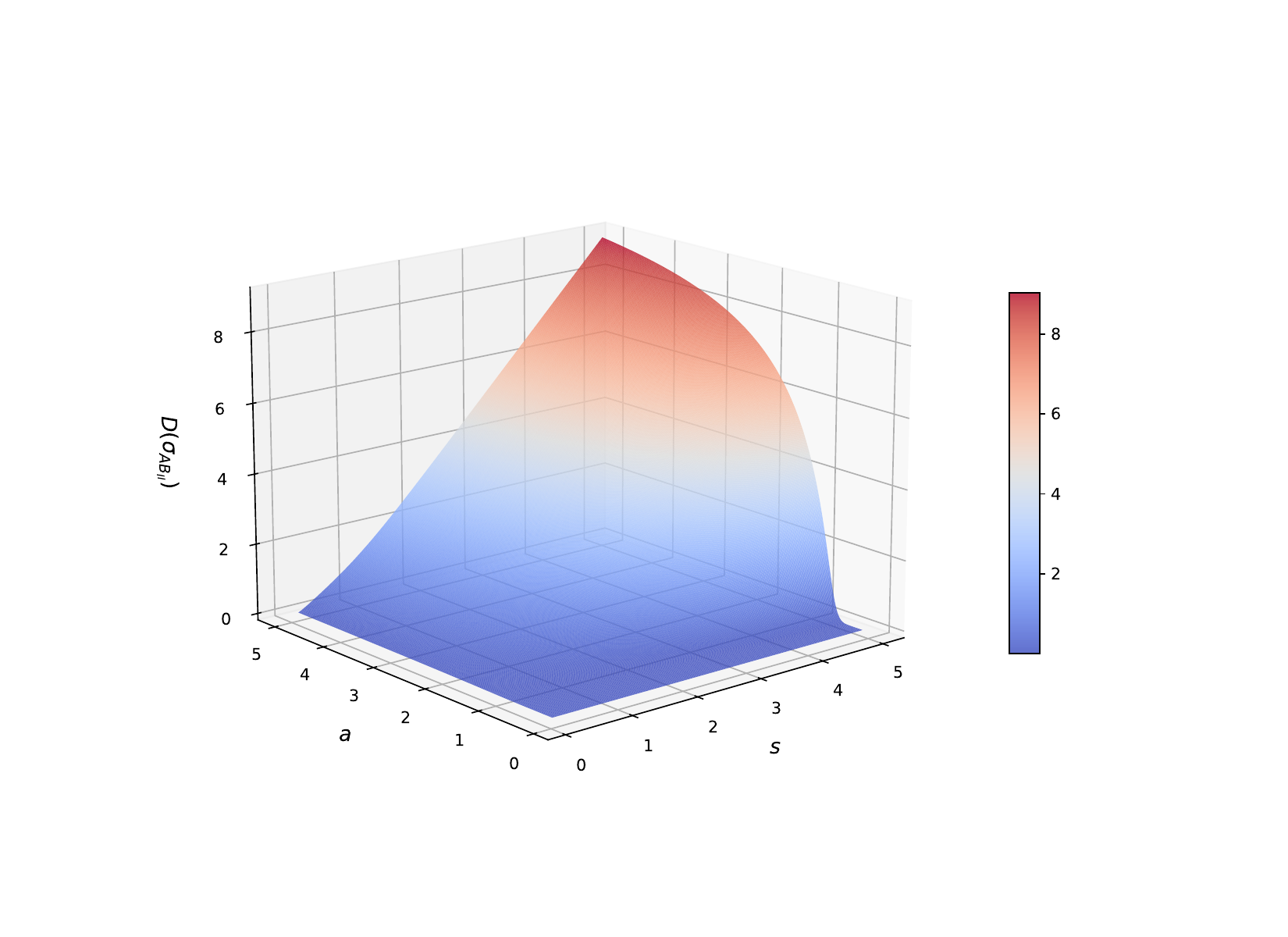}
		\\[-3.5em]  
		\hspace*{-2.7em}(a)
	\end{minipage}
	\hfill
	\begin{minipage}[b]{0.7\columnwidth}
		\centering
		\includegraphics[width=\linewidth]{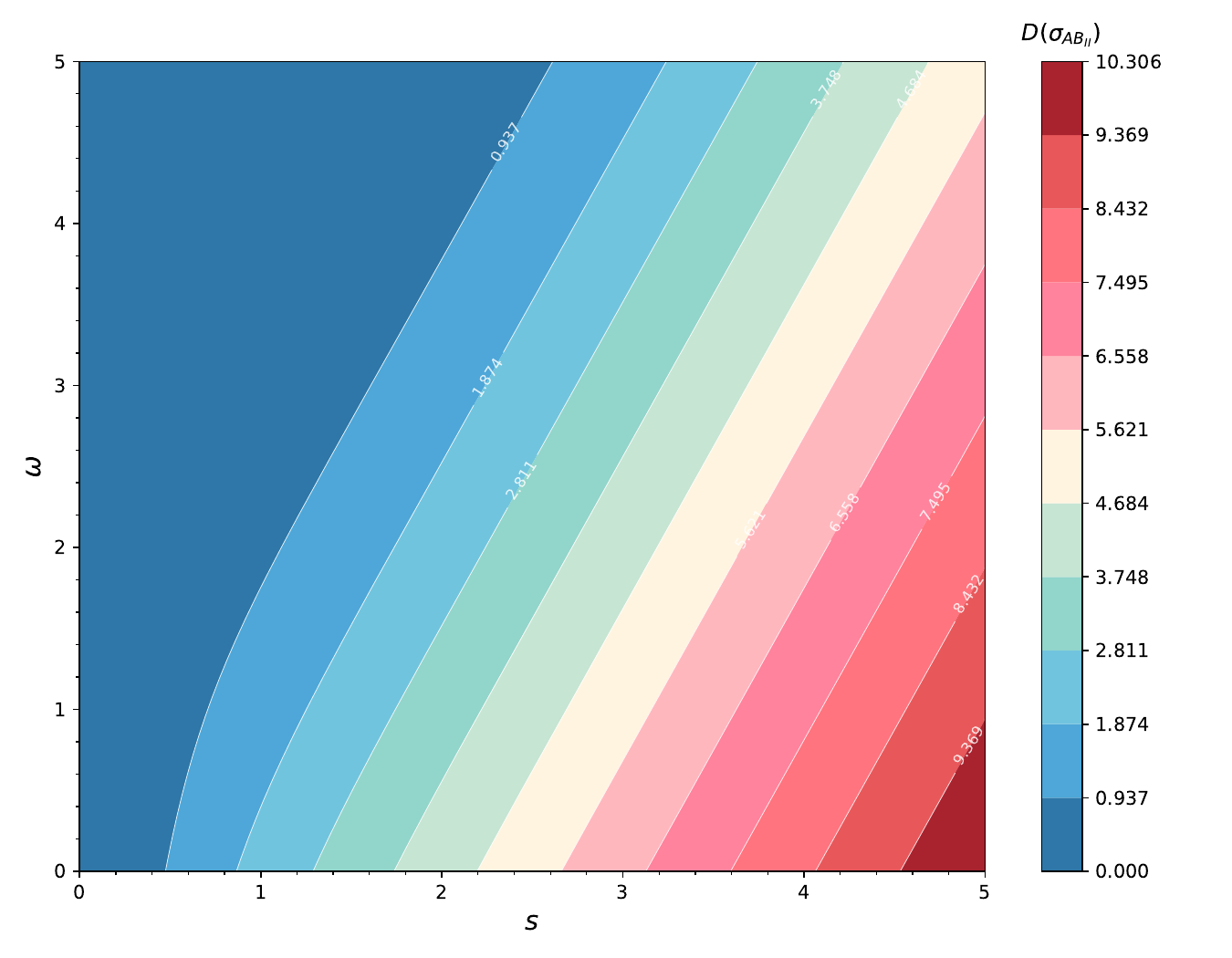}
		\\[-0.5em]
		\hspace*{-2.7em}(b)
	\end{minipage}
	\caption{Gaussian quantum discord of the mode pair $A$-$B_{II}$. (a) Discord as a function of the acceleration $a$ and the squeezing parameter $s$ for a fixed field frequency $\omega=1$. (b) Discord as a function of the field frequency $\omega$ and the squeezing parameter $s$ for a fixed acceleration $a=2\pi$.}
	\label{fig:3D-ABII}
\end{figure}

We now consider Bob moving with a uniform acceleration $a$, such that the state corresponding to mode $k$ must be described in Rindler coordinates (see Fig.~\ref{fig:Bjiasu}), so that the Minkowski vacuum state takes the form
\begin{equation}
	\ket{0_k}_{M} = U^{I,II}_k \ket{0_k}_{I}\ket{0_k}_{II}.
\end{equation}
Namely, as Bob undergoes uniform acceleration, the state description from his perspective incorporates an additional two-mode squeezing transformation, where the squeezing parameter $r_k$ is related to Bob's acceleration $a$ via Eq.~\eqref{eqr}.

\begin{figure}[t]
	\centering
	\begin{minipage}[b]{0.7\columnwidth}
		\centering
		\includegraphics[width=\linewidth]{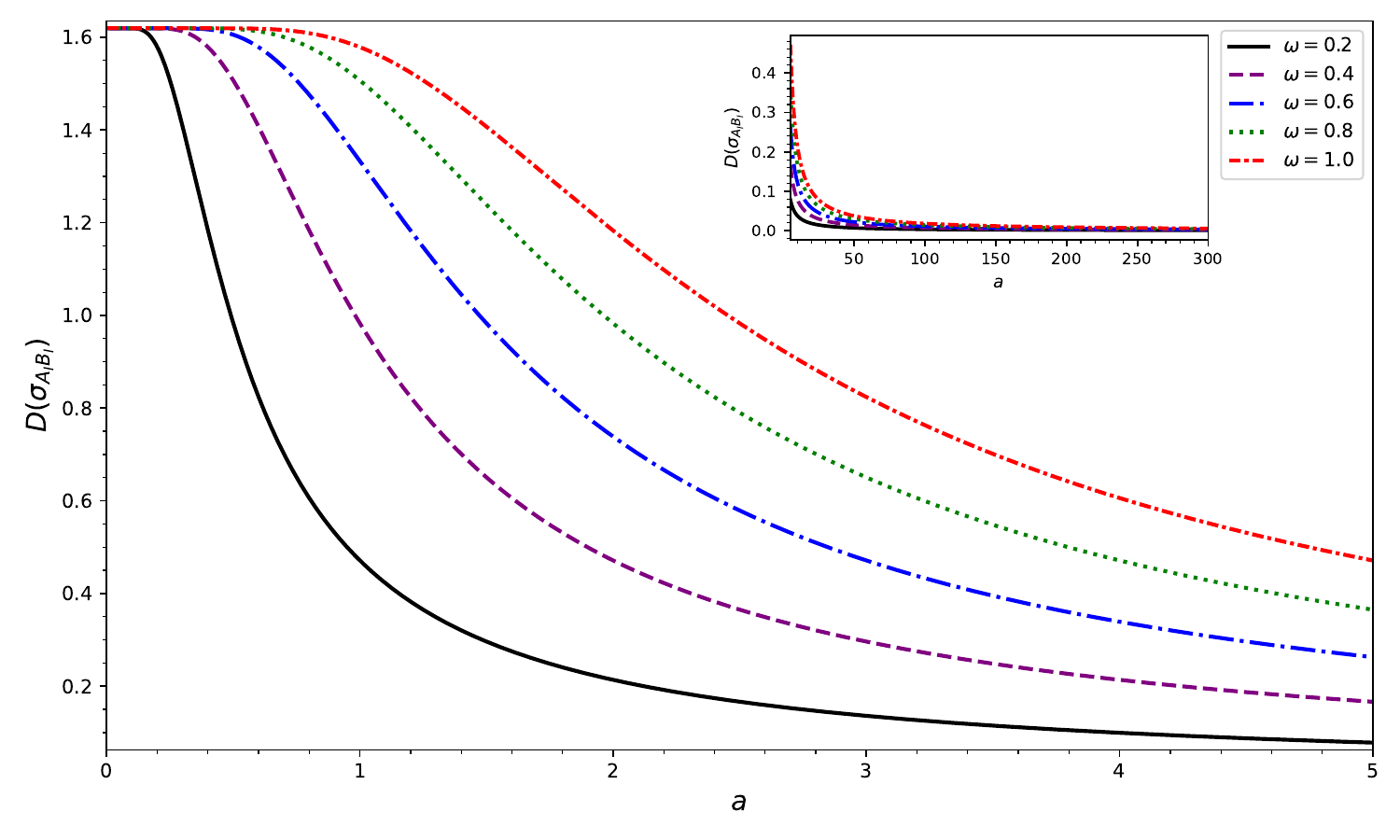}
		\\[-0.5em]  
		\hspace*{-1.5em}(a)
	\end{minipage}
	\hfill
	\begin{minipage}[b]{0.7\columnwidth}
		\centering
		\includegraphics[width=\linewidth]{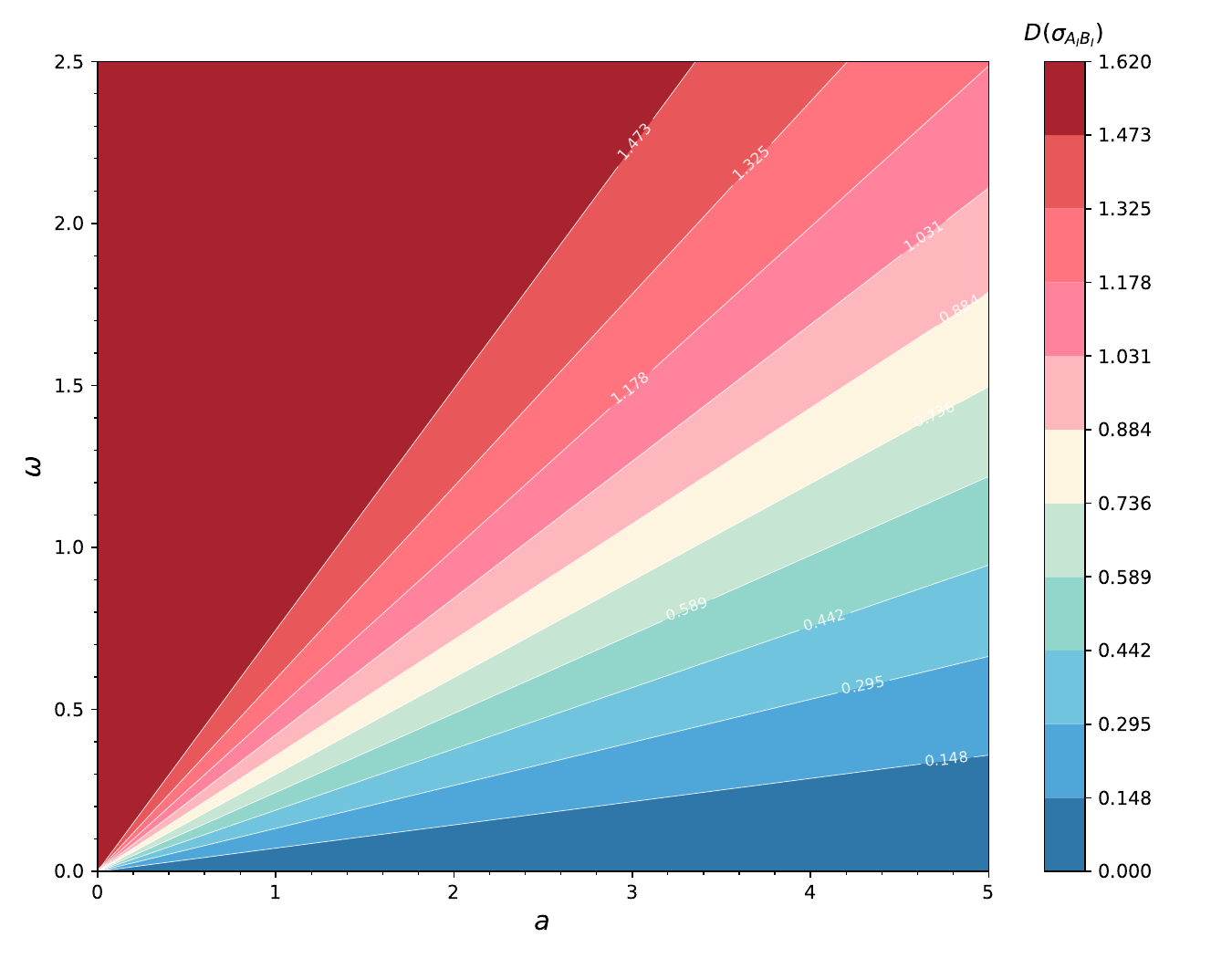}
		\\[-0.5em]
		\hspace*{-2.5em}(b)
	\end{minipage}
	\caption{Gaussian quantum discord of the mode pair $A_I$-$B_{I}$ as a function of the acceleration $a$ and the field frequency $\omega$, with the squeezing parameter fixed at $s=1$.}
	\label{fig:2D-AIBI}
\end{figure}

Upon the two-mode squeezing transformation, Bob's mode $B$ is mapped onto a pair of modes: $B_I$ corresponding to the accelerated Bob in Rindler region I, and $B_{II}$ corresponding to the virtual anti-Bob in the causally disconnected Rindler region $II$. The entire system is thus partitioned into three subsystems: subsystem $A$ described by the inertial observer Alice, subsystem $B_I$ described by Bob in Rindler region $I$, and subsystem $B_{II}$ described by the virtual observer anti-Bob. The total covariance matrix of this tripartite system after the transformation reads as follows:
\begin{equation}
	\sigma_{AB_{I}B_{II}} = \bigl[I_A \oplus S_{B_{I}B_{II}}\bigr] \bigl[\sigma_{AB} \oplus I_{B_{II}}\bigr] \bigl[I_A \oplus S_{B_{I}B_{II}}\bigr]^{\mathrm{T}},
	\label{eq:tripartite_cov_transform}
\end{equation}
where $I_A \oplus S_{B_{I}B_{II}}$ describes the two-mode squeezing transformation between subsystems $B_{I}$ and $B_{II}$, and $\sigma_{AB} \oplus I_{B_{II}}$ denotes the initial covariance matrix of the total system.

Since Rindler regions $I$ and $II$ are causally disconnected, Bob confined to Rindler region I has no access to the mode $B_{II}$ in region $II$. Therefore, the reduced covariance matrix of the Alice–Bob subsystem can be obtained by tracing over the $B_{II}$ mode.
\begin{equation}
	\sigma_{AB_{I}} = \begin{pmatrix}
		\mathcal{A}_{AB_{I}} & \mathcal{C}_{AB_{I}} \\
		\mathcal{C}_{AB_{I}}^\mathrm{T} & \mathcal{B}_{AB_{I}}
	\end{pmatrix},
	\label{eq:ABI}
\end{equation}
where
\begin{equation}
	\begin{split}
		\mathcal{A}_{AB_{I}} &= \frac{\cosh(2s)}{2}\, I_2, \\
		\mathcal{C}_{AB_{I}} &= \frac{\sinh(2s)}{2 \sqrt{1 - \theta^2_k}} \, Z_2, \\
		\mathcal{B}_{AB_{I}} &= \left[\frac{\cosh(2s) + \theta_k^2}{2-2\theta_k^2}\right]\, I_2.\notag
	\end{split}
\end{equation}

The reduced covariance matrices for the Bob–anti-Bob subsystem and the Alice–anti-Bob subsystem can be obtained by tracing out mode $A$ and mode $B_I$ from the total covariance matrix, respectively, i.e.,
\begin{equation}
	\sigma_{B_{{I}} B_{II}} = \begin{pmatrix}
		\mathcal{A}_{B_{{I}} B_{{II}}} & \mathcal{C}_{B_{{I}} B_{{II}}} \\
		\mathcal{C}_{B_{{I}} B_{{II}}}^\mathrm{T} & \mathcal{B}_{B_{{I}} B_{{II}}}
	\end{pmatrix},
	\label{eq:BIBII}
\end{equation}
with elements
\begin{equation*}
	\begin{aligned}
		\mathcal{A}_{B_{{I}} B_{{II}}} &= \left[\frac{\cosh(2s) + \theta_k^2}{2-2\theta_k^2}\right]\, I_2, \\
		\mathcal{B}_{B_{{I}} B_{{II}}} &= \left[\frac{\theta^2_k \cosh(2s) + 1}{2 - 2 \theta^2_k}\right]\, I_2, \\
		\mathcal{C}_{B_{{I}} B_{{II}}} &= \frac{\theta_k \cosh^2(s)}{1 - \theta^2_k}\, Z_2.
	\end{aligned}
\end{equation*}

And \begin{equation}
	\sigma_{A B_{{II}}} = \begin{pmatrix}
		\mathcal{A}_{A B_{{II}}} & \mathcal{C}_{A B_{{II}}} \\
		\mathcal{C}_{A B_{{II}}}^\mathrm{T} & \mathcal{B}_{A B_{{II}}}
	\end{pmatrix},
	\label{eq:ABII}
\end{equation}
where
\begin{equation*}
	\begin{aligned}
		\mathcal{A}_{A B_{{II}}} &= \frac{\cosh(2s)}{2}\, I_2, \\
		\mathcal{B}_{A B_{{II}}} &= \left[\frac{\cosh(2s) + \theta_k^2}{2-2\theta_k^2}\right]\, I_2, \\
		\mathcal{C}_{A B_{{II}}} &= \frac{\sinh(2s)}{2 \sqrt{1 - \theta^2_k}}\, Z_2.
	\end{aligned}
\end{equation*}

Using the covariance matrices given by~\eqref{eq:ABI}, \eqref{eq:BIBII} and \eqref{eq:ABII}, together with Eq.~\eqref{eq:quantum_discord_sts}, we analyze the influence of the Unruh effect on Gaussian quantum discord between inertial and non-inertial observers.

In Fig.~\ref{fig:2D-ABI-BIBII-ABII}, we show the variation of Gaussian quantum discord with the acceleration $a$ and the field frequency $\omega$ for different mode pairs. We observe that:

\noindent (i) For the accessible mode pair $A$-$B_{I}$, the quantum discord decreases monotonically as the acceleration $a$ grows, and drops to zero in the infinite-acceleration limit $a\to\infty$. The acceleration $a$ and the field frequency $\omega$ exhibit competing influences on the Gaussian quantum discord: an elevated field frequency $\omega$ effectively suppresses the Unruh-induced decay of quantum discord, whereas a lower $\frac{\omega}{a}$ ratio accelerates the degradation of the quantum discord.

\noindent (ii) For both the $B_{I}$-$B_{II}$ and $A$-$B_{II}$ mode pairs, the quantum discord vanishes in the inertial limit ($a=0$) and grows monotonically with increasing acceleration. Specifically, the quantum discord of the $A$-$B_{II}$ mode pair gradually saturates to a frequency-independent maximum, whereas that of the $B_{I}$-$B_{II}$ mode pair remains unbounded. The acceleration $a$ and the field frequency $\omega$ also exert competing influences on the Gaussian quantum discord: an elevated field frequency $\omega$ effectively suppresses the Unruh-driven growth of quantum discord, while a lower $\frac{\omega}{a}$ ratio enhances this growth.

\noindent (iii) The Unruh effect does not completely eliminate the initial Gaussian quantum discord; instead, it redistributes the quantum discord across the Rindler subsystems. In particular, the degradation of discord between the inertial mode and the accessible noninertial mode in region $I$ is compensated by the emergence of quantum discord in two causally disconnected mode pairs: the inertial mode paired with the region $II$ mode, and the region $I$ accessible noninertial mode paired with the region $II$ mode.

In Figs.~\ref{fig:3D-ABI}, \ref{fig:3D-BIBII}, and \ref{fig:3D-ABII}, we respectively present the Gaussian quantum discord for different mode pairs as a function of the acceleration $a$ and the squeezing parameter $s$, as well as the dependence of discord on the field frequency $\omega$ and the squeezing parameter $s$. We find that:

\noindent (i) For an arbitrary squeezing parameter $s$ and fixed field frequency $\omega$, the dependence of Gaussian quantum discord on acceleration $a$ for each mode pair follows the same qualitative behavior as presented in Fig.~\ref{fig:2D-ABI-BIBII-ABII}.

\noindent (ii) With the field frequency $\omega$ fixed, a smaller squeezing parameter $s$ reduces the sensitivity of quantum discord to acceleration, endowing the quantum correlation with stronger robustness against the Unruh effect.

\begin{figure*}
	\centering
	\subfloat[$A_I$-$B_{II}$]{\includegraphics[width=0.32\textwidth]{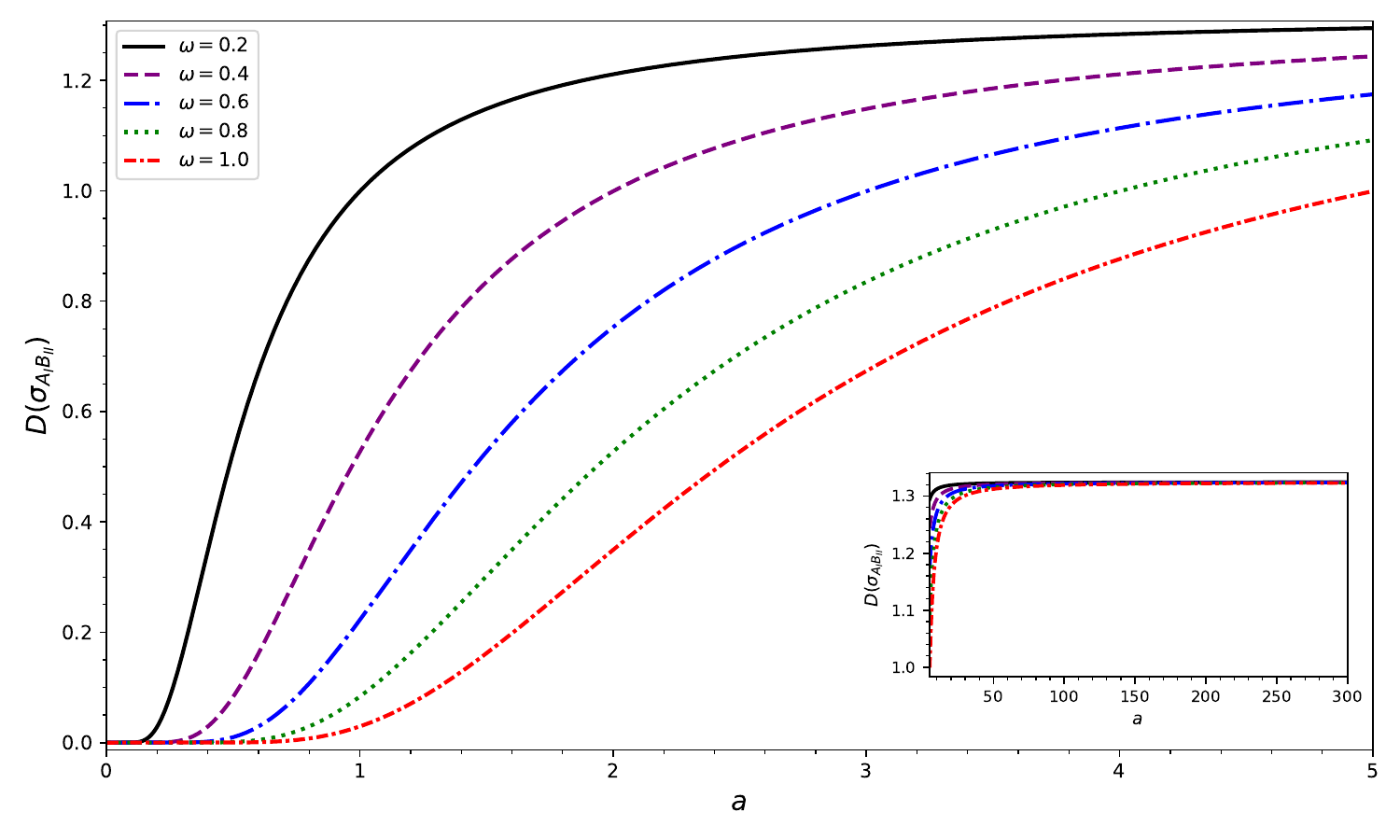}}
	\hfill
	\subfloat[$A_{II}$-$B_{II}$]{\includegraphics[width=0.32\textwidth]{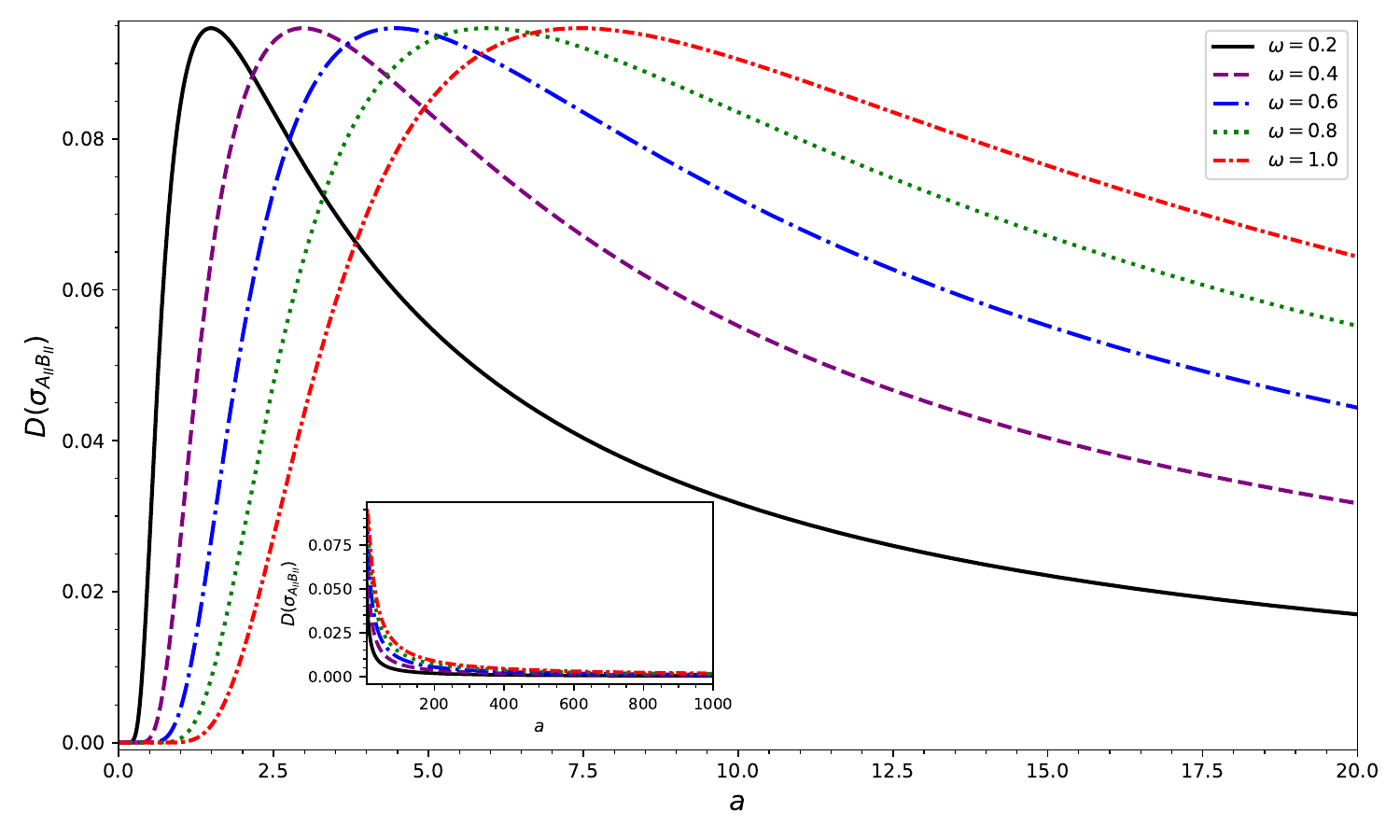}}
	\hfill
	\subfloat[$A_I$-$A_{II}$]{\includegraphics[width=0.32\textwidth]{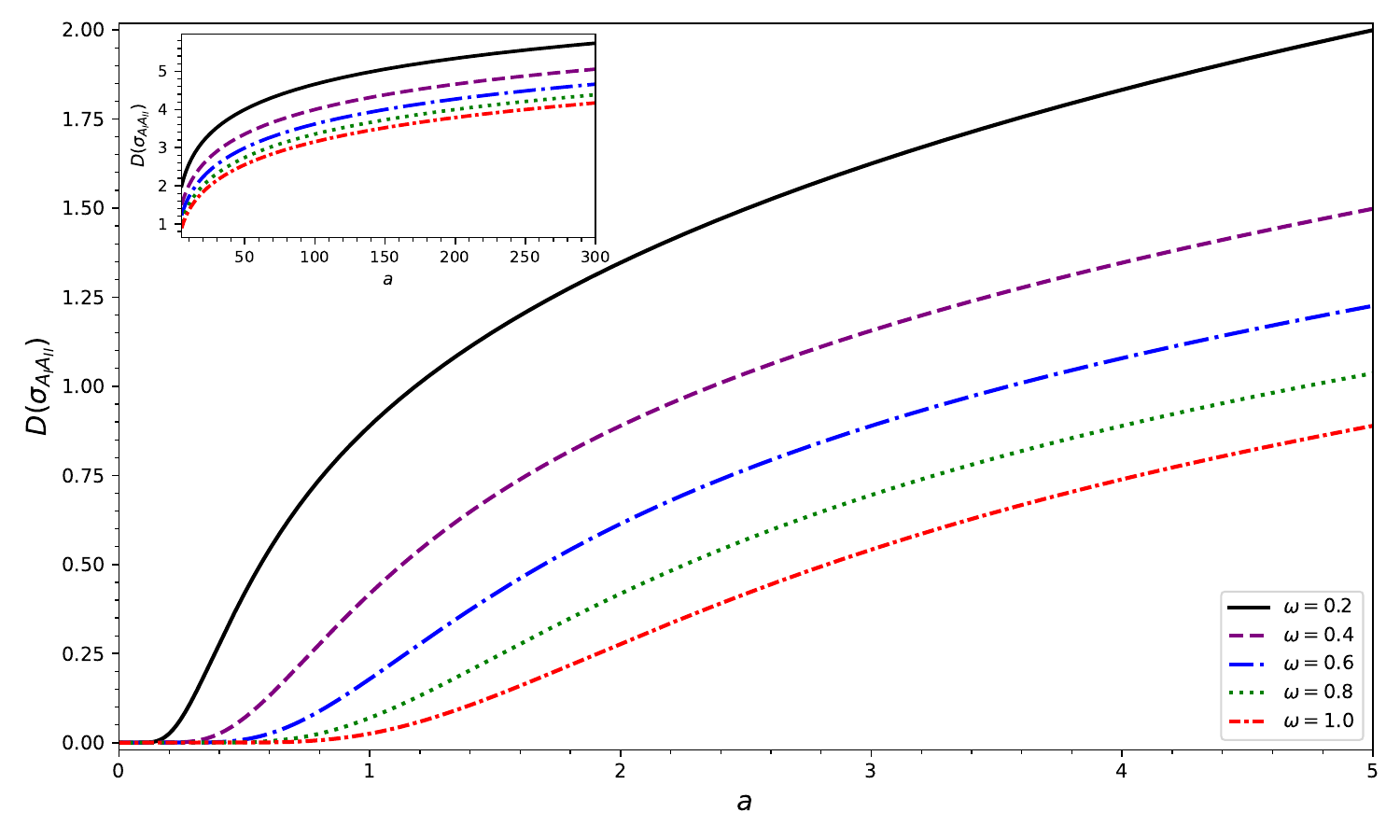}}
	
	\vspace{0em} 
	
	\subfloat[$A_I$-$B_{II}$]{\includegraphics[width=0.32\textwidth]{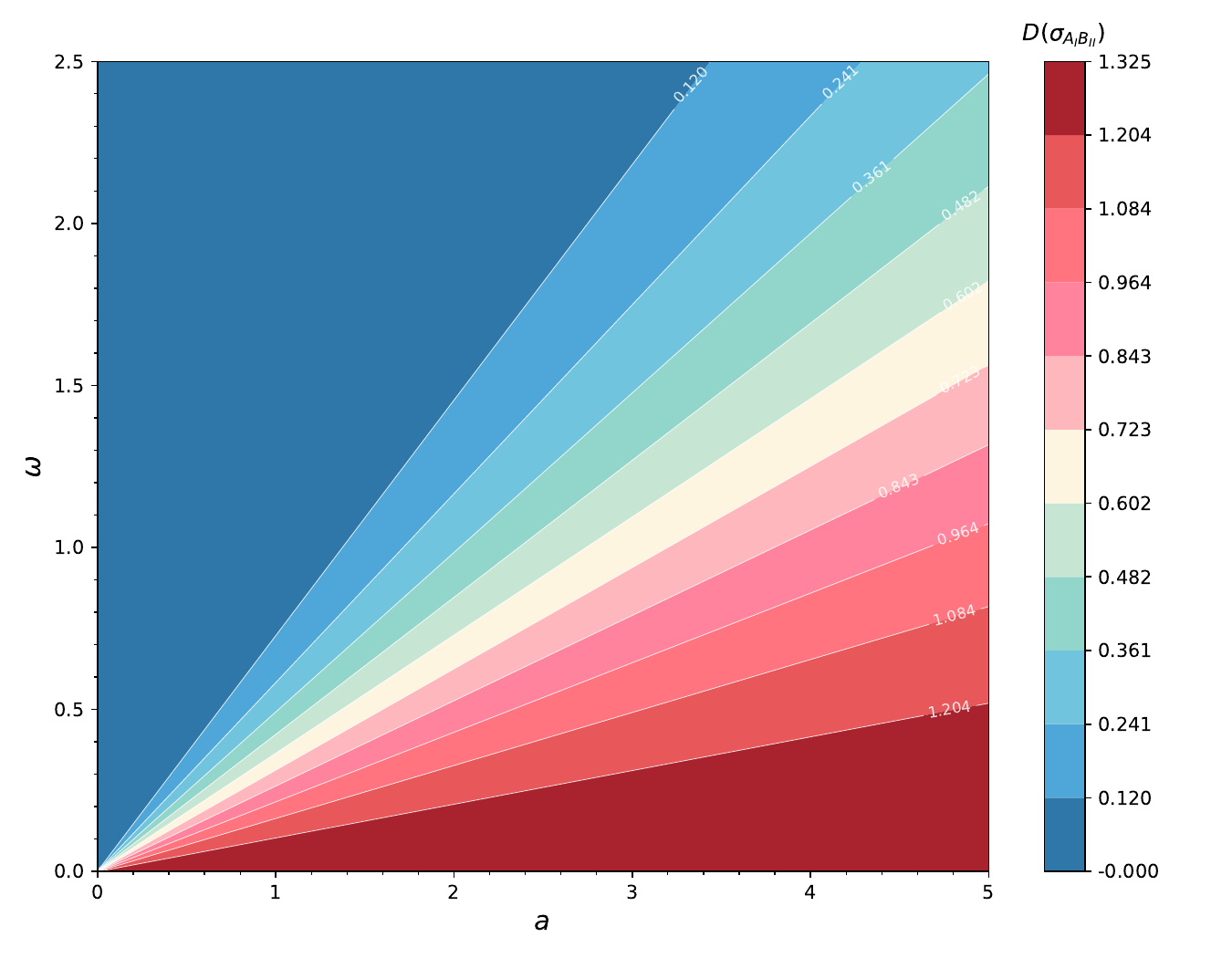}}
	\hfill
	\subfloat[$A_{II}$-$B_{II}$]{\includegraphics[width=0.32\textwidth]{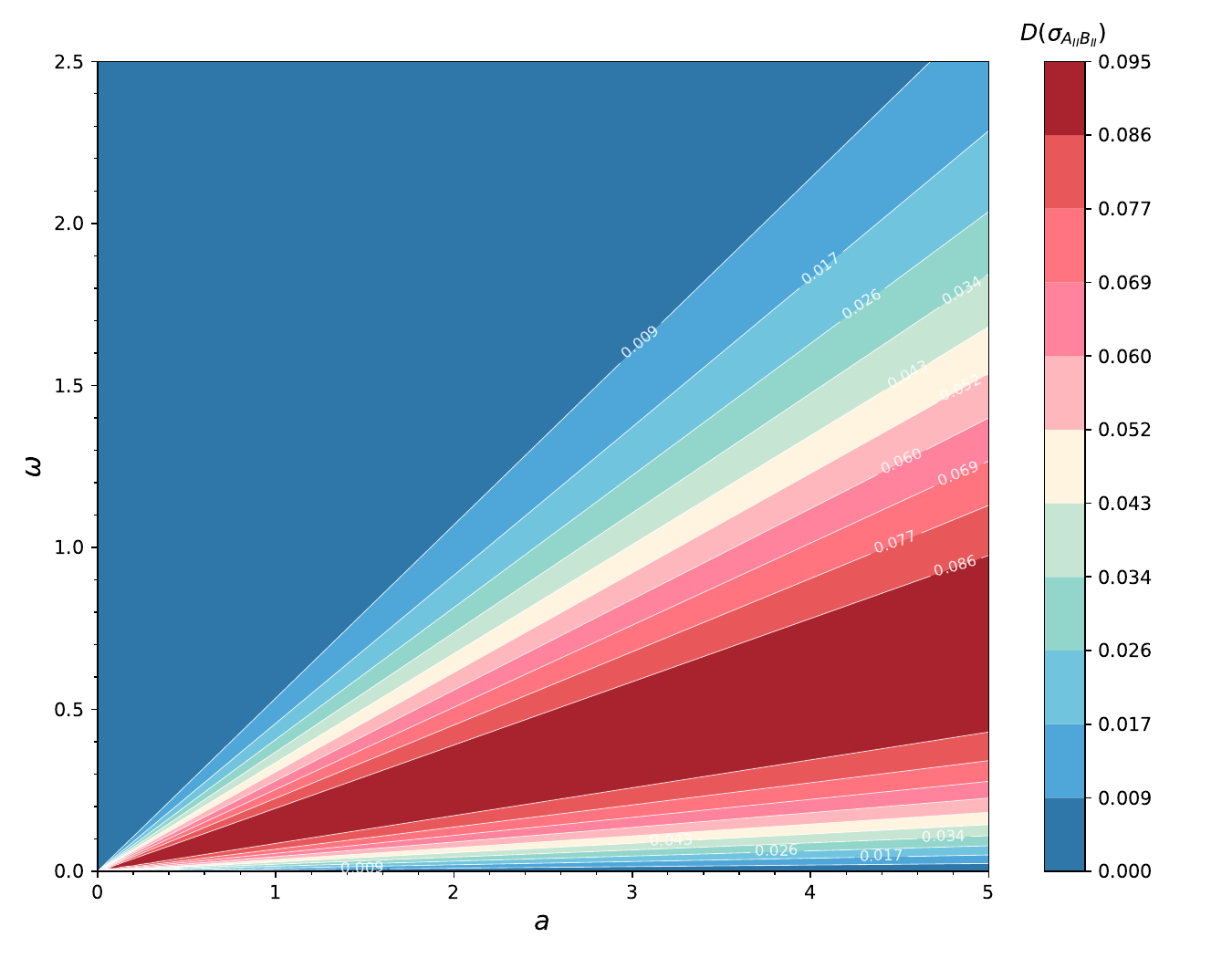}}
	\hfill
	\subfloat[$A_I$-$A_{II}$]{\includegraphics[width=0.32\textwidth]{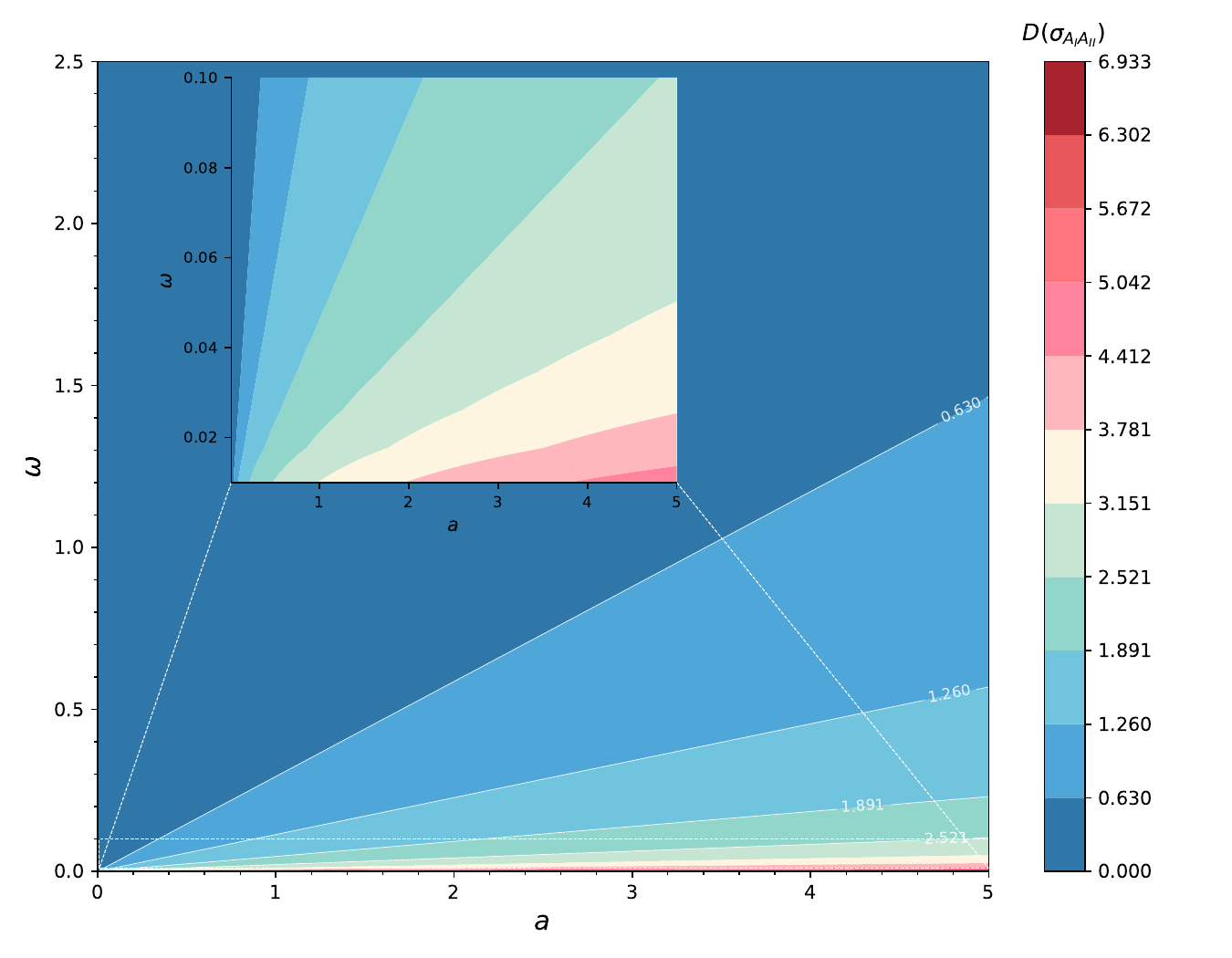}}
	
	\caption{Gaussian quantum discord for different mode pairs as a function of the acceleration $a$ and the field frequency $\omega$, with the squeezing parameter fixed at $s=1$.}
	\label{fig:2D-AIBII-AIIBII-AIAII}
\end{figure*}

\noindent (iii) With the acceleration $a$ fixed, the dependence of Gaussian quantum discord on $s$ and $\omega$ exhibits distinct features for different mode pairs: (1) For the $A$-$B_{I}$ mode pair, the squeezing parameter $s$ and the field frequency $\omega$ have comparable influences on the discord: $D$ grows monotonically with both $\omega$ (at fixed $s$) and $s$ (at fixed $\omega$). (2) For the $A$-$B_{II}$ mode pair, the discord remains negligible for $s<\omega$ (e.g., $s=0.5$, $\omega=3$) even at very low frequencies, whereas it increases rapidly to values above 6 once $s>\omega$ (e.g., $s=4$, $\omega=2$), revealing that the squeezing parameter $s$ plays the dominant role in this case. (3) For the $B_{I}$-$B_{II}$ mode pair, remarkably, the discord plummets as soon as $\omega$ departs slightly from zero, whereas the squeezing parameter $s$ only has a weak modulating effect on $D$ across the entire interval $[0,5]$. This indicates that the field frequency $\omega$ governs the behavior of discord in an overwhelming manner.

\subsection{Gaussian quantum discord distribution induced by two uniformly accelerated observers}
Does the quantum discord redistribution mechanism induced by the Unruh effect undergo a qualitative change with the number of uniformly accelerated observers? We now consider the setting where both Alice and Bob undergo uniform acceleration $a$, and characterize the state of a scalar field from their noninertial frame of reference, formulated in terms of Rindler coordinates (see Fig.~\ref{fig:ABjiasu}).

Analogously, the two-mode squeezing transformation maps each of the initial modes $A$ (Alice) and $B$ (Bob) onto a pair of Rindler modes: $A_{I}$ and $B_{I}$ pertain to the accelerated observers Alice and Bob in Rindler region $I$, while $A_{II}$ and $B_{II}$ are associated with the fictitious anti-Alice and anti-Bob residing in the causally disconnected Rindler region $II$.

\begin{figure}
	\centering
	\begin{minipage}[b]{0.9\columnwidth}
		\centering
		\includegraphics[width=\linewidth,trim={10 0 0 100},clip]{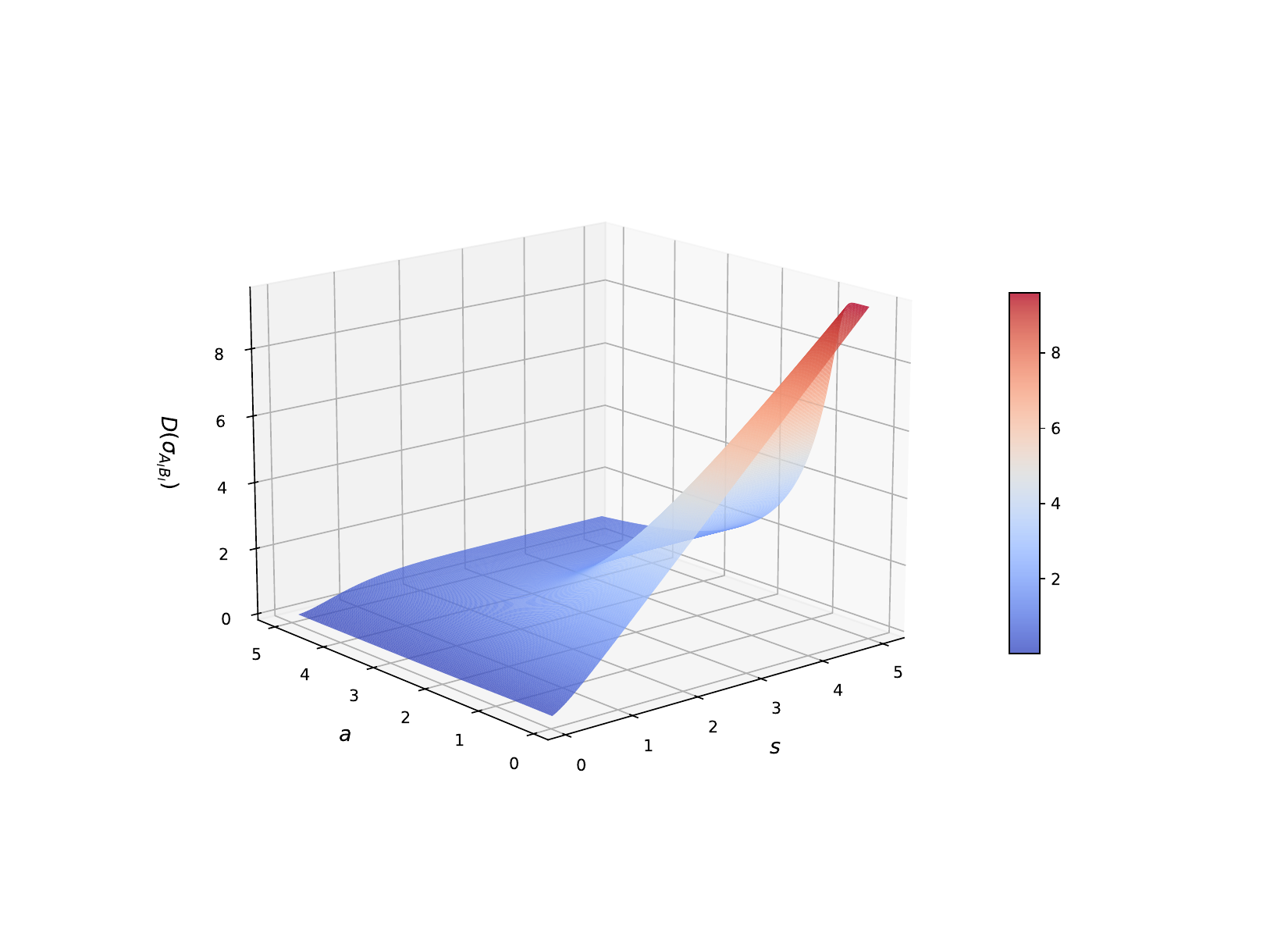}
		\\[-3.5em]  
		\hspace*{-2.6em}(a)
	\end{minipage}
	\hfill
	\begin{minipage}[b]{0.7\columnwidth}
		\centering
		\includegraphics[width=\linewidth]{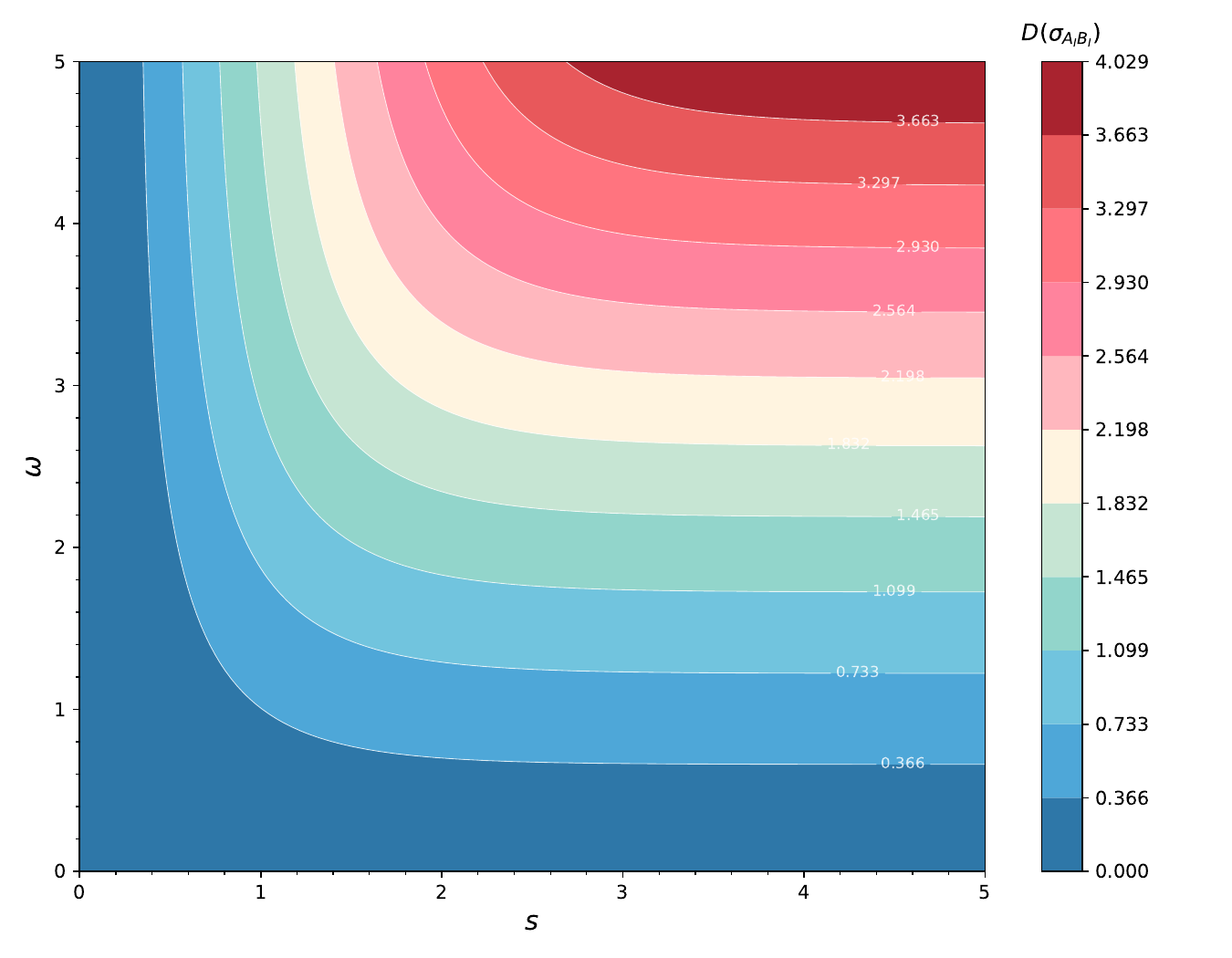}
		\\[-0.5em]
		\hspace*{-2.6em}(b)
	\end{minipage}
	\caption{Gaussian quantum discord of the mode pair $A_I$-$B_{I}$. (a) Discord as a function of the acceleration $a$ and the squeezing parameter $s$ for a fixed field frequency $\omega=1$. (b) Discord as a function of the field frequency $\omega$ and the squeezing parameter $s$ for a fixed acceleration $a=2\pi$.}
	\label{fig:3D-AIBI}
\end{figure}

The full system is thus divided into four subsystems: subsystems $A_{I}$ and $B_{I}$ described by Alice and Bob in Rindler region $I$, and subsystems $A_{II}$ and $B_{II}$ described by the fictitious observers anti-Alice and anti-Bob. The total covariance matrix of this four-partite system after the transformation is given by:
\begin{equation}
	\begin{split}
			\sigma_{A_IB_IA_{II}B_{II}} =
			&\left[ S_{A_I,A_{II}} \oplus S_{B_I,B_{II}} \right] \left[ \sigma_{AB} \oplus I_{A_{II},B_{II}} \right] \\
			&\left[ S_{A_I,A_{II}} \oplus S_{B_I,B_{II}} \right]^\mathrm{T},
		\label{eq:4mode_cov_evolution}
	\end{split}
\end{equation}
where $S_{A_{I},A_{II}} \oplus S_{B_{I},B_{II}}$ describes the two-mode squeezing transformation between subsystems $A_{I}$--$A_{II}$ and $B_{I}$--$B_{II}$, while $\sigma_{AB} \oplus I_{A_{II},B_{II}}$ denotes the initial covariance matrix of the full system.

\begin{figure}
	\centering
	\includegraphics[width=0.8\linewidth]{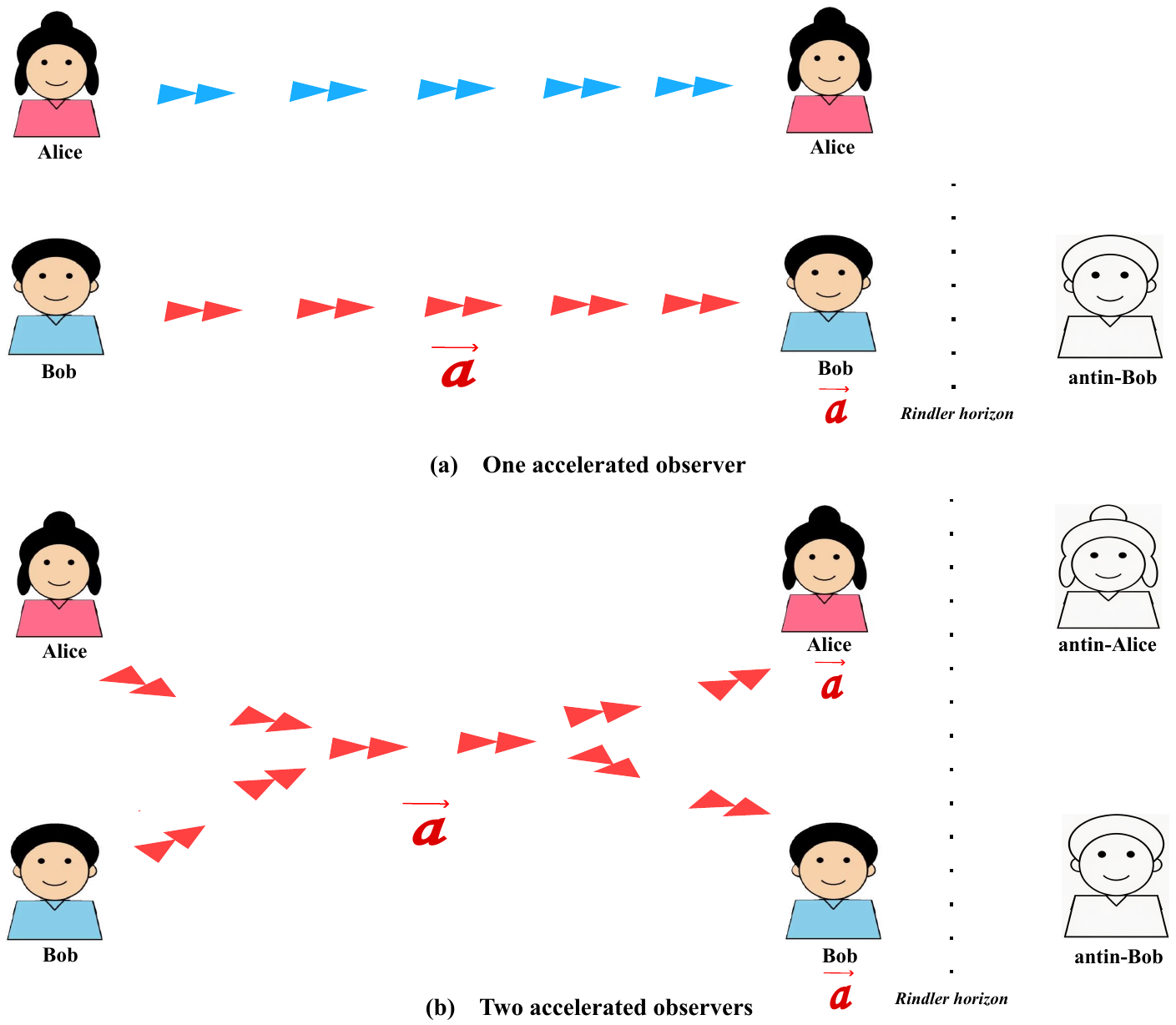}
	\caption{(a) The case where Bob undergoes uniform acceleration. (b) The case where both Alice and Bob undergo uniform acceleration.
	}
	\label{fig:shiyitu}
\end{figure}

As Rindler regions $I$ and $II$ are causally disconnected, Alice and Bob restricted to region $I$ cannot probe the unobserved modes $A_{II}$ and $B_{II}$ in the causally isolated region $II$. We accordingly trace out these inaccessible degrees of freedom, yielding the reduced covariance matrix for the Alice--Bob subsystem:
\begin{equation}
	\sigma_{A_IB_I} = \begin{pmatrix}
		\mathcal{A}_{A_IB_I} & \mathcal{C}_{A_IB_I} \\
		\mathcal{C}_{A_IB_I}^\mathrm{T} & \mathcal{B}_{A_IB_I}
	\end{pmatrix},
	\label{eq:A_IB_I}
\end{equation}
where
\begin{equation}
	\begin{split}
		\mathcal{A}_{A_IB_I} &= 	\left[\frac{\cosh(2s) + \theta_k^2}{2-2\theta_k^2}\right]\, I_2, \\
		\mathcal{C}_{A_IB_I} &= \frac{\sinh(2s)}{2-2\theta_k^2} \,Z_2, \\
		\mathcal{B}_{A_IB_I} &= 	\left[\frac{\cosh(2s) + \theta_k^2}{2-2\theta_k^2}\right]\, I_2.\notag
	\end{split}
\end{equation}

Next, we trace out modes $A_I$ and $B_I$ from the total covariance matrix to obtain the reduced covariance matrix of the anti-Alice--anti-Bob subsystem, which reads:
\begin{equation}
	\sigma_{A_{II}B_{II}} = \begin{pmatrix}
		\mathcal{A}_{A_{II}B_{II}} & \mathcal{C}_{A_{II}B_{II}} \\
		\mathcal{C}_{A_{II}B_{II}}^\mathrm{T} & \mathcal{B}_{A_{II}B_{II}}
	\end{pmatrix},
	\label{eq:A_IIB_II}
\end{equation}
where
\begin{equation}
	\begin{split}
		\mathcal{A}_{A_{II}B_{II}} &=\left[\dfrac{\theta_k^2 \cosh(2s) + 1}{2 - 2\theta_k^2}\right]\, I_2, \\
		\mathcal{C}_{A_{II}B_{II}} &= \dfrac{\theta_k^2 \sinh(2s)}{2 - 2\theta_k^2}\, Z_2, \\
		\mathcal{B}_{A_{II}B_{II}} &=\left[\dfrac{\theta_k^2 \cosh(2s) + 1}{2 - 2\theta_k^2}\right]\, I_2.\notag
	\end{split}
\end{equation}

Secondly, by tracing out modes $B_I$ and $A_{II}$, and modes $A_I$ and $B_{II}$, respectively, we obtain the covariance matrices $\sigma_{A_IB_{II}}$ (Alice-anti-Bob) and $\sigma_{A_{II}B_I}$ (anti-Alice-Bob).
\begin{equation}
	\sigma_{A_IB_{II}} = \sigma_{A_{II}B_I} = \begin{pmatrix}
		\mathcal{A}_{A_{II}B_I} & \mathcal{C}_{A_{II}B_I} \\
		\mathcal{C}_{A_{II}B_I}^\mathrm{T} & \mathcal{B}_{A_{II}B_I}
	\end{pmatrix},
	\label{eq:A_IB_II}
\end{equation}
where
\begin{equation}
	\begin{split}
		\mathcal{A}_{A_{II}B_I} &=\left[\dfrac{\cosh(2s) + \theta_k^2}{2 - 2\theta_k^2}\right]\, I_2, \\
		\mathcal{C}_{A_{II}B_I} &= \dfrac{\theta_k \sinh(2s)}{2 - 2\theta_k^2}\, I_2, \\
		\mathcal{B}_{A_{II}B_I} &=\left[\dfrac{\theta_k^2 \cosh(2s) + 1}{2 - 2\theta_k^2}\right]\, I_2.\notag
	\end{split}
\end{equation}

Finally, we turn our attention to the quantum discord between modes $A_I$ and $A_{II}$, and between modes $B_I$ and $B_{II}$. By tracing out modes $B_I$ and $B_{II}$ and modes $A_I$ and $A_{II}$, respectively, we obtain the covariance matrices for Alice-anti-Alice and Bob-anti-Bob, i.e., $\sigma_{A_IA_{II}}$ and $\sigma_{B_IB_{II}}$.
\begin{equation}
	\sigma_{A_IA_{II}} = \sigma_{B_IB_{II}} = \begin{pmatrix}
		\mathcal{A}_{B_IB_{II}} & \mathcal{C}_{B_IB_{II}} \\
		\mathcal{C}_{B_IB_{II}}^\mathrm{T} & \mathcal{B}_{B_IB_{II}}
	\end{pmatrix},
	\label{eq:A_IA_II}
\end{equation}
where
\begin{equation}
	\begin{split}
		\mathcal{A}_{B_IB_{II}} &=\left[\dfrac{\cosh(2s) + \theta_k^2}{2 - 2\theta_k^2}\right]\, I_2, \\
		\mathcal{C}_{B_IB_{II}} &= \dfrac{2\theta_k \cosh^2(s)}{2 - 2\theta_k^2}\, Z_2, \\
		\mathcal{B}_{B_IB_{II}} &=\left[\dfrac{\theta_k^2 \cosh(2s) + 1}{2 - 2\theta_k^2}\right]\, I_2.\notag
	\end{split}
\end{equation}

Similarly, using the covariance matrices given by~\eqref{eq:A_IB_I}, \eqref{eq:A_IIB_II}, \eqref{eq:A_IB_II} and~\eqref{eq:A_IA_II}, combined with the expressions in Eqs.~\eqref{eq:gaussian_discordyiban} and~\eqref{eq:quantum_discord_sts}, we analyze the influence of the Unruh effect on the Gaussian quantum discord between two noninertial observers.

\begin{figure*}
	\centering
	\subfloat[$A_I$-$B_{II}$]{\includegraphics[width=0.32\textwidth]{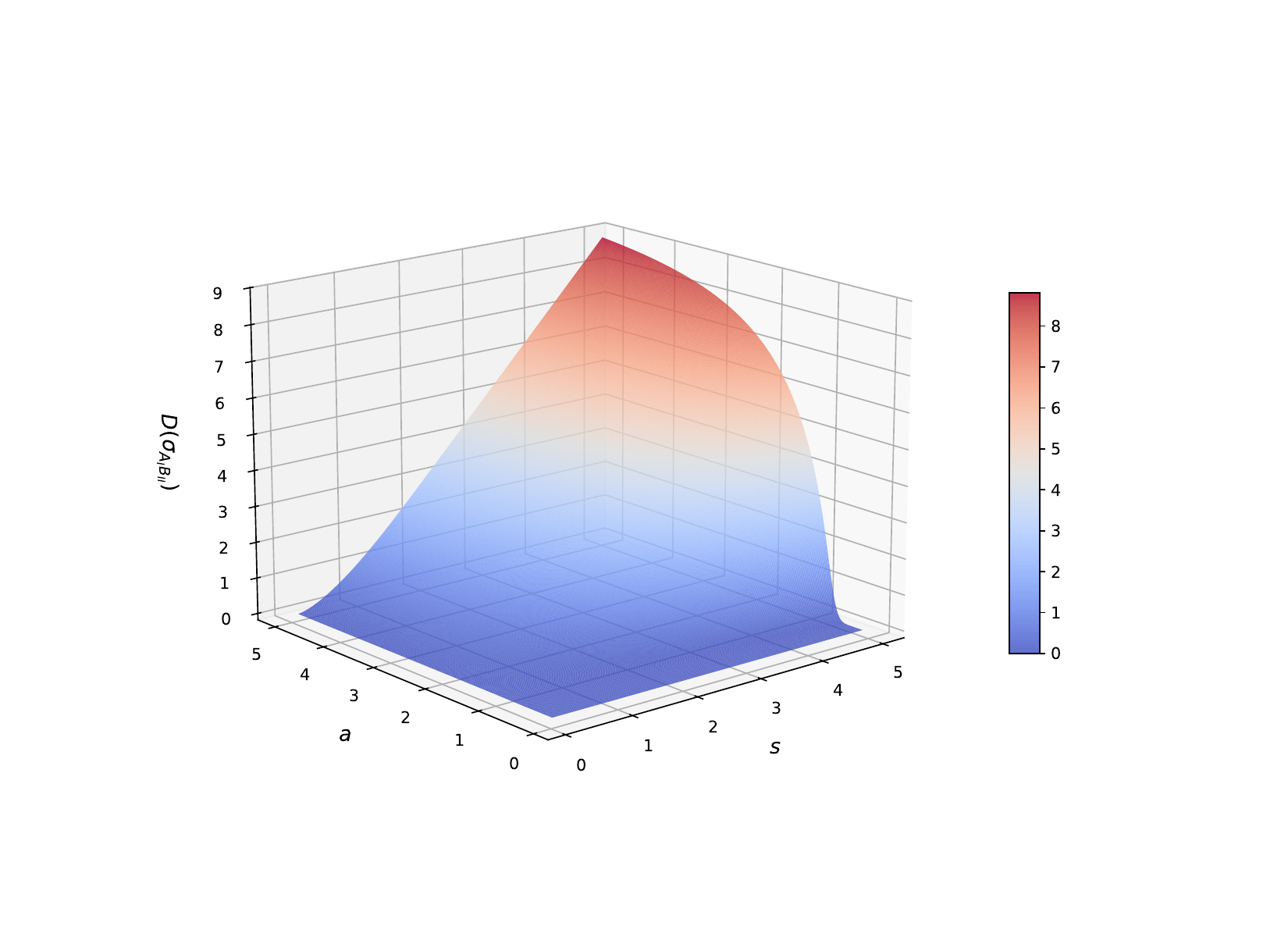}}
	\hfill
	\subfloat[$A_{II}$-$B_{II}$]{\includegraphics[width=0.32\textwidth]{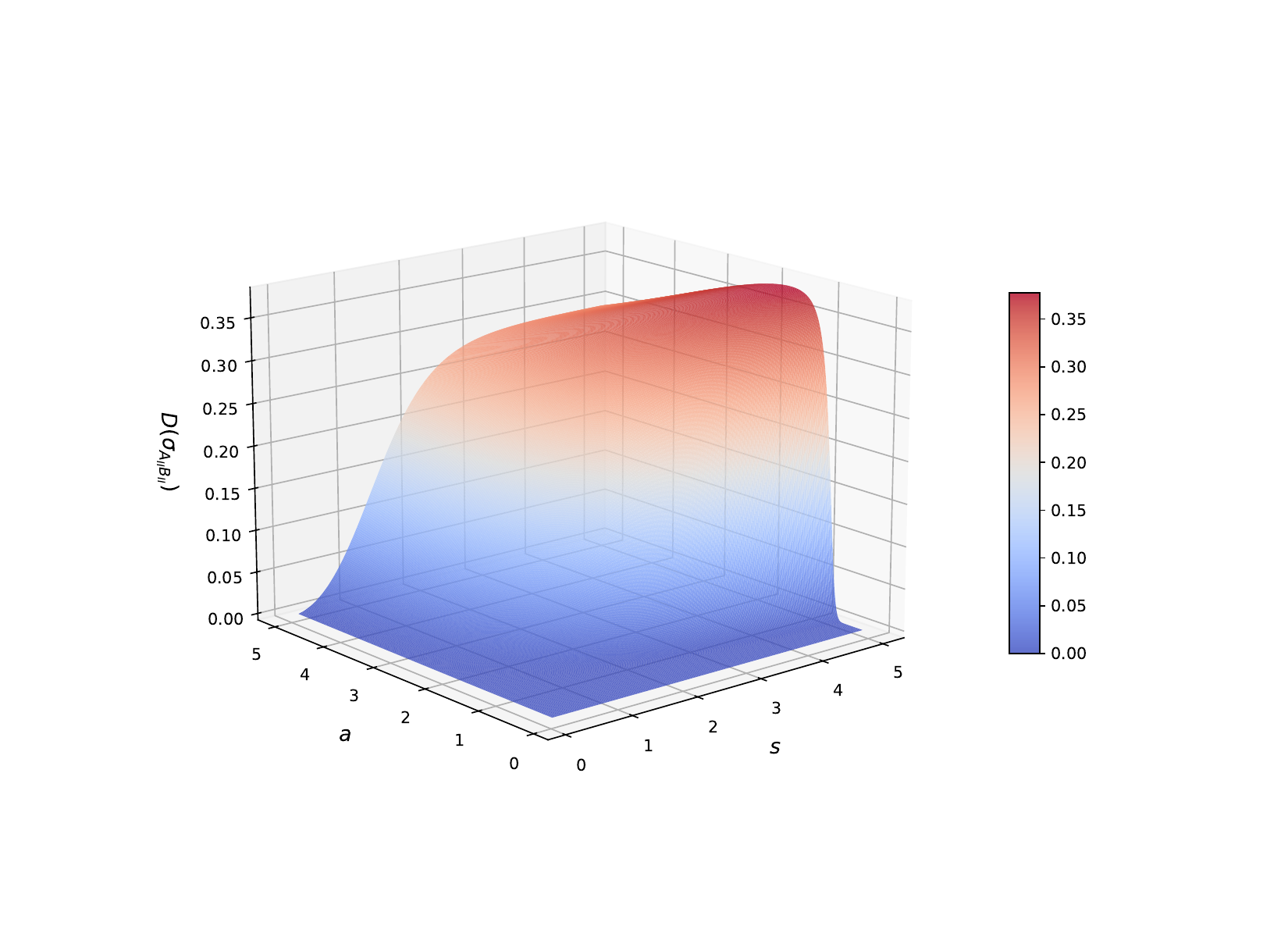}}
	\hfill
	\subfloat[$A_I$-$A_{II}$]{\includegraphics[width=0.32\textwidth]{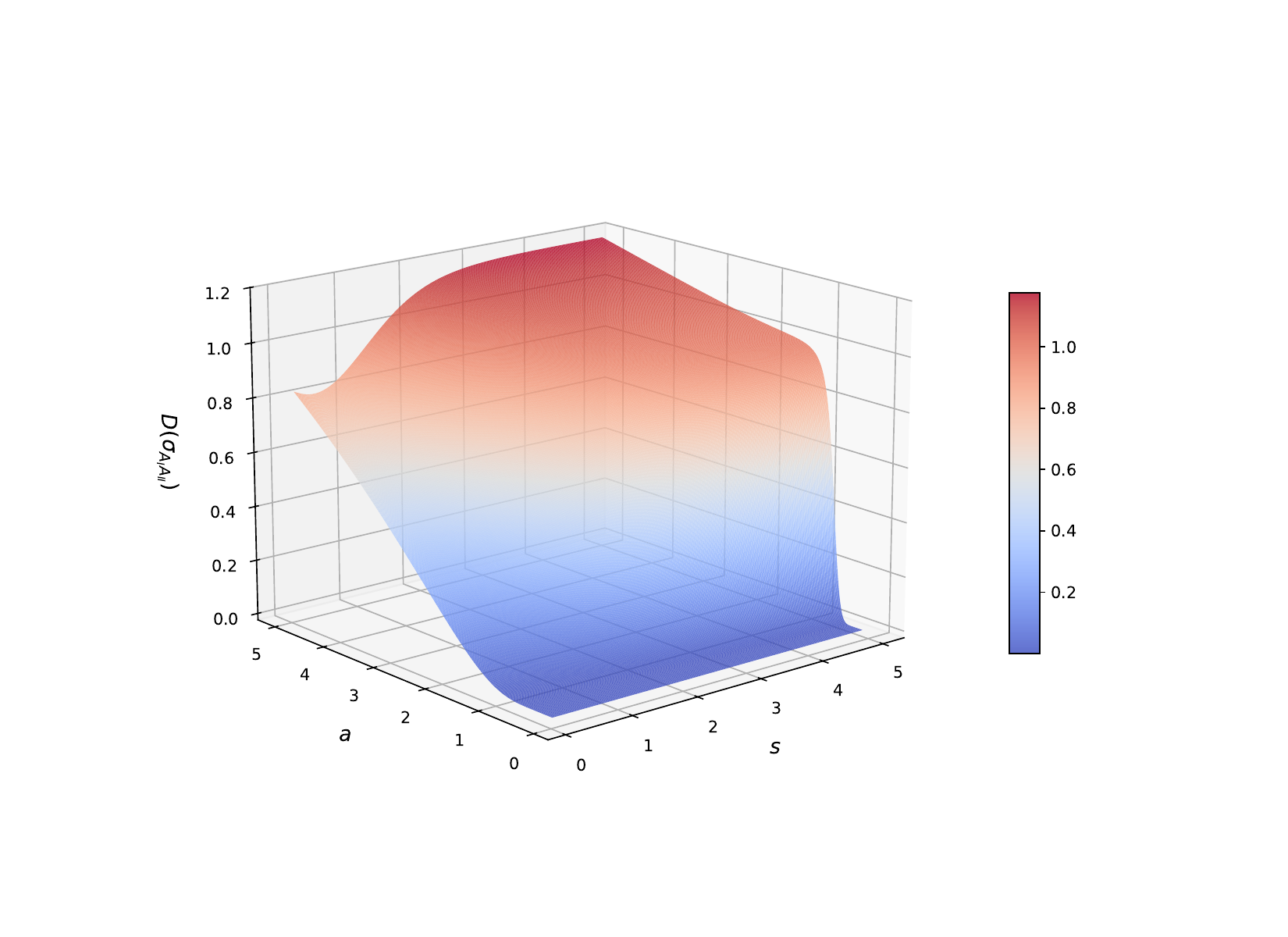}}
	
	\vspace{0em} 
	
	\subfloat[$A_I$-$B_{II}$]{\includegraphics[width=0.32\textwidth]{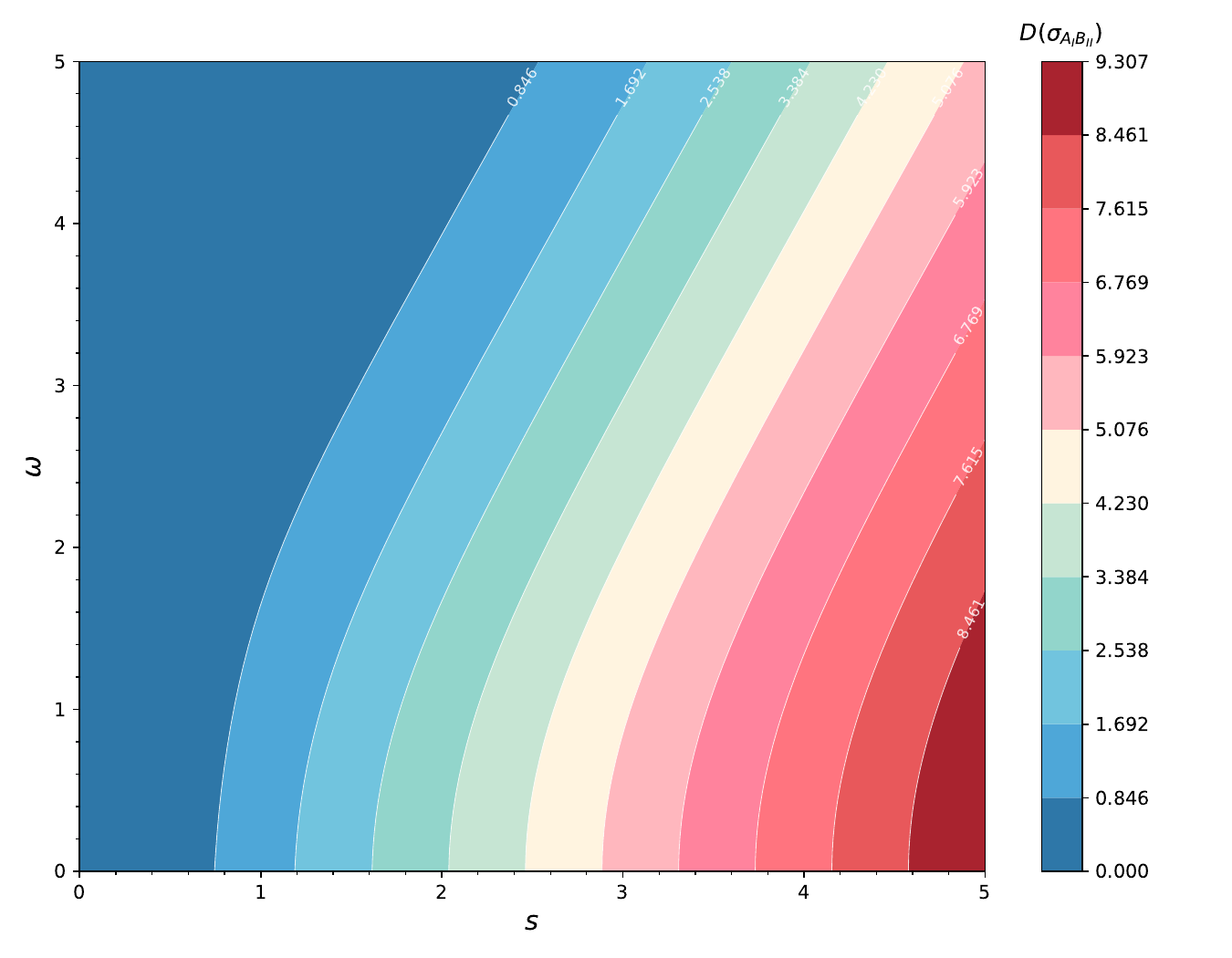}}
	\hfill
	\subfloat[$A_{II}$-$B_{II}$]{\includegraphics[width=0.32\textwidth]{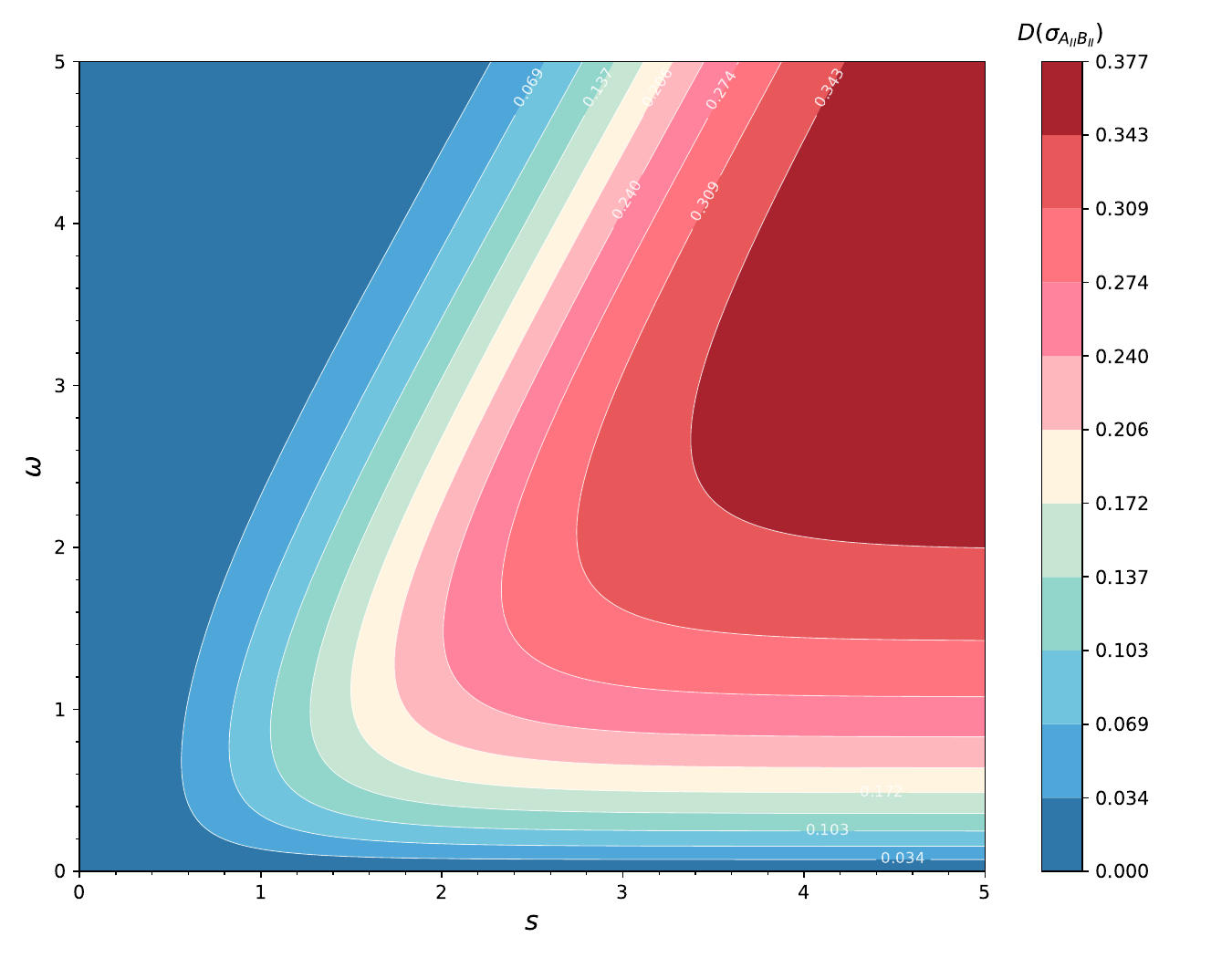}}
	\hfill
	\subfloat[$A_I$-$A_{II}$]{\includegraphics[width=0.32\textwidth]{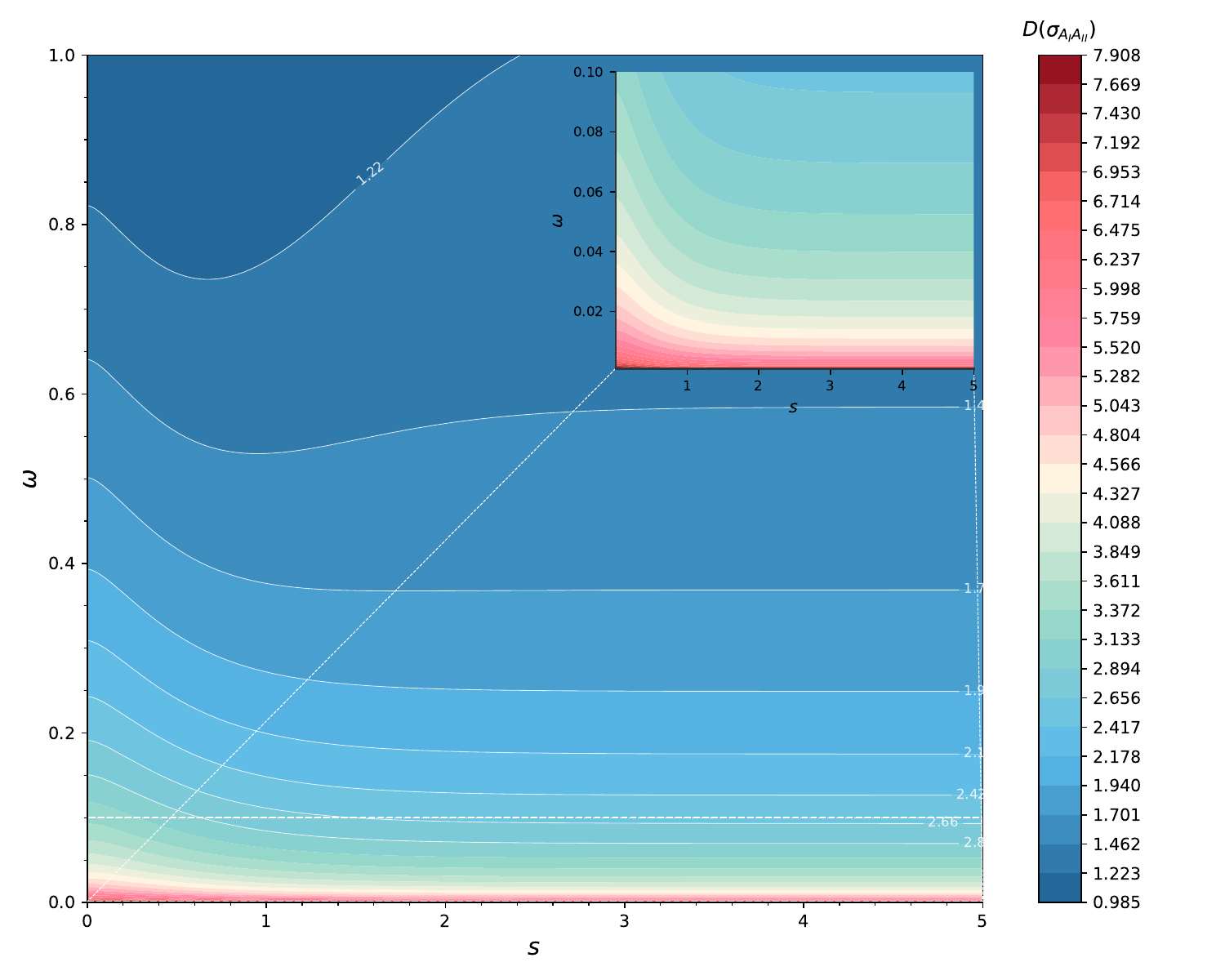}}
	
	\caption{Gaussian quantum discord for different mode pairs. (a)--(c) Discord as a function of the acceleration $a$ and the squeezing parameter $s$ with the field frequency fixed at $\omega=1$. (d)--(f) Discord as a function of the field frequency $\omega$ and the squeezing parameter $s$ with the acceleration fixed at $a=2\pi$.}
	\label{fig:3D-AIBII-AIIBII-AIAII}
\end{figure*}

In Figs.~\ref{fig:2D-AIBI} and \ref{fig:2D-AIBII-AIIBII-AIAII}, we show the variation of Gaussian quantum discord with the acceleration $a$ and the field frequency $\omega$ for different mode pairs. Interestingly, this behavior exhibits a qualitative difference compared with the scenario involving only one accelerated observer:
\noindent (i) For the accessible mode pair $A_{I}$-$B_{I}$, the Gaussian quantum discord decays monotonically with acceleration $a$ and vanishes in the infinite-acceleration limit $a\to\infty$. Consistently, the acceleration $a$ and the field frequency $\omega$ impose competing influences on the discord: an elevated field frequency $\omega$ effectively suppresses the Unruh-induced degradation of quantum correlation, whereas a lower $\frac{\omega}{a}$ ratio expedites the decay of discord.
\noindent (ii)  For both $A_{I}$-$B_{II}$ and $A_{I}$-$A_{II}$ mode pairs, the quantum discord is absent in the inertial limit ($a=0$) and increases monotonically with acceleration. In detail, the discord of the $A_{I}$-$B_{II}$ pair saturates to a frequency-independent maximum, while that of the $A_{I}$-$A_{II}$ pair remains unbounded. The two parameters again exhibit competing effects: a higher field frequency $\omega$ mitigates the Unruh-driven growth of discord, and a smaller $\frac{\omega}{a}$ ratio enhances this growth.
\noindent (iii) The $A_{II}$--$B_{II}$ mode pair shows markedly different behavior: (1) Its quantum discord first increases to a peak value with growing acceleration, then declines, and eventually tends to zero as $a\to\infty$. Furthermore, the acceleration corresponding to the maximum discord shifts to larger values as the field frequency $\omega$ increases. (2) The growth of quantum discord depends non-monotonically on the ratio $\frac{\omega}{a}$: it is most pronounced at an optimal $\frac{\omega}{a}$ value, and becomes markedly weaker when the ratio is either too large or too small.
\noindent (iv) The Gaussian quantum discord exhibits a distinct hierarchical distribution: the intra-observer Rindler $I$-$II$ mode pairs of each single accelerated observer carry the largest discord, followed by the inter-observer cross-region $I$-$II$ mode pairs, while the inter-observer mode pair with both modes in the causally disconnected region $II$ has the lowest discord.

In Figs.~\ref{fig:3D-AIBI} and \ref{fig:3D-AIBII-AIIBII-AIAII}, we respectively present the Gaussian quantum discord for different mode pairs as a function of the acceleration $a$ and the squeezing parameter $s$, as well as the dependence of discord on the field frequency $\omega$ and the squeezing parameter $s$. We find that:

\noindent (i) For an arbitrary squeezing parameter $s$ and fixed field frequency $\omega$, the dependence of Gaussian quantum discord on acceleration $a$ for each mode pair follows the same qualitative behavior as presented in Fig.~\ref{fig:2D-AIBII-AIIBII-AIAII}. With the field frequency $\omega$ fixed, a smaller squeezing parameter $s$ reduces the sensitivity of quantum discord to acceleration, endowing the quantum correlation with stronger robustness against the Unruh effect.

\noindent (ii) For a fixed acceleration $a$, the dependence of Gaussian quantum discord on $s$ and $\omega$ exhibits highly diverse features across different mode pairs: 
(1) For the accessible $A_{I}$-$B_{I}$ mode pair, the squeezing parameter $s$ and the field frequency $\omega$ contribute comparably to the quantum discord.
(2) For the $A_{I}$-$B_{II}$ cross-region mode pair, the squeezing parameter $s$ remains the dominant factor governing the discord magnitude.
(3) For the $A_{II}$-$B_{II}$ mode pair entirely contained in Rindler region $II$, the contour lines adopt a hyperbolic profile, yet the discord magnitude is strictly confined below 0.4 over the full parameter space. This implies that adjusting $s$ or $\omega$ cannot substantially enhance the discord inside region II, where both parameters exert similarly weak effects. The discord becomes non-negligible only when the $\frac{\omega}{s}$ ratio falls into an optimal interval.
(4) The behavior of the $A_{I}$-$A_{II}$ mode pair is qualitatively identical to that observed in the single-accelerated-observer scenario. 

\section{conclusions}

In this work, we investigate the redistribution of continuous-variable Gaussian quantum discord under the Unruh effect and examine how the key system parameters influence the quantum correlation. We find that an appropriately chosen field frequency, together with a suitable squeezing parameter, facilitates the extraction of information about the Unruh effect.

We further reveal the redistribution law of the initial quantum discord. For both the single- and two-accelerated-observer scenarios, the discord between the mode pair between Alice and Bob decays monotonically with increasing acceleration and vanishes in the infinite-acceleration limit $a\to\infty$. Concurrently, the Unruh effect induces quantum discord in all causally disconnected mode pairs, namely $B_{I}$-$B_{II}$, $A$-$B_{II}$, $A_{I}$-$B_{II}$ (or $B_{I}$-$A_{II}$), $A_{II}$-$B_{II}$, and $A_{I}$-$A_{II}$. This confirms that the initial quantum discord is redistributed, rather than destroyed, across the different Rindler partitions.

By comparing the single- and two-observer cases, we identify distinct parametric roles for different classes of mode pairs. For the causally connected pairs $A$-$B_{I}$ and $A_{I}$-$B_{I}$, the squeezing parameter $s$ and the field frequency $\omega$ exert comparable and mutually interchangeable effects on the discord. For the cross-region pairs $A$-$B_{II}$ and $A_{I}$-$B_{II}$, the squeezing parameter $s$ plays a decisive role in determining the discord magnitude. For the intra-observer cross-region pairs $A_{I}$-$A_{II}$ and $B_{I}$-$B_{II}$, the field frequency $\omega$ exerts a more prominent influence.

In the two-observer scenario, the $A_{II}$-$B_{II}$ mode pair exhibits distinctive non-monotonic behavior: its discord first rises to a peak and then declines with increasing acceleration, with the peak position shifting to larger $a$ as the field frequency $\omega$ increases. In this case, both the squeezing parameter $s$ and the field frequency $\omega$ contribute similarly weak effects to the discord, which becomes non-negligible only when the ratio $\frac{\omega}{s}$ falls within an optimal interval.

Our work significantly enriches the theoretical framework of Gaussian quantum discord under the Unruh effect, and provides new perspectives and a solid theoretical foundation for subsequent research.

\begin{acknowledgments}
This work is supported by the Natural Science Foundation of Hainan Province under Grant No. 125RC744; the China Scholarship Council (CSC).
\end{acknowledgments}

\bibliography{ref}

\end{document}